\documentclass[12pt]{article}
\usepackage{jheppub}
\usepackage{amssymb,amsfonts}

\usepackage{wrapfig}
\usepackage{epsfig}
\usepackage[T1]{fontenc}
\usepackage[utf8]{inputenc}
\usepackage{lmodern}
\usepackage{float}
\usepackage{placeins}
\usepackage{dsfont}
\usepackage{arydshln}
\usepackage{bm}
\usepackage{bbold}

\usepackage[dvipsnames]{xcolor}
\usepackage{tikz}

\def\be{\begin{eqnarray}}
\def\ee{\end{eqnarray}}
\newcommand{\nn}{\nonumber}
\newcommand\para{\paragraph{}}
\newcommand{\ft}[2]{{\textstyle\frac{#1}{#2}}}
\newcommand{\eqn}[1]{(\ref{#1})}

\newcommand{\ppp}[2]{\frac{\partial {#1}}{\partial {#2}}}

\newcommand{\blambda}{\bm{\lambda}}

\newcommand{\bphi}{\bm{\phi}}
\newcommand{\bpi}{\bm{\pi}}
\newcommand{\bnu}{\bm{\bnu}}

\newcommand{\bsigma}{\bm{\sigma}}

\newcommand{\btheta}{\bm{\theta}}

\newcommand{\ra}{\rightarrow}
\newcommand{\ket}[1]{|#1\rangle}
\newcommand{\bra}[1]{\langle#1|}

\def\Dslash{\,\,{\raise.15ex\hbox{/}\mkern-12mu D}}
\def\Dbarslash{\,\,{\raise.15ex\hbox{/}\mkern-12mu {\bar D}}}
\def\delslash{\,\,{\raise.15ex\hbox{/}\mkern-9mu \partial}}
\def\delbarslash{\,\,{\raise.15ex\hbox{/}\mkern-9mu {\bar\partial}}}
\def\pslash{\,\,{\raise.15ex\hbox{/}\mkern-9mu p}}
\def\calDslash{\,\,{\raise.15ex\hbox{/}\mkern-12mu {\cal D}}}

\newcommand{\Z}{{\mathds Z}}

\newcommand{\R}{{\mathds R}}

\newcommand{\br}{{\bf r}}

\def\lae{\mathrel{\mathop{\smash{\lower .5 ex \hbox{$\stackrel<\sim$}}}}}
\def\lae{\mathrel{\mathop{\smash{\lower .5 ex \hbox{$\stackrel>\sim$}}}}}

\newcommand{\tchi}{\tilde{\chi}}

\title{What Comes Out Of Monopole-Fermion Scattering?}

\author[1,2]{David Tong}

\affiliation[1]{Department of Applied Mathematics and Theoretical Physics \\ University of Cambridge, CB3 0WA, UK}
\affiliation[2]{CERN, Theoretical Physics Department, Geneva, Switzerland}

\emailAdd{d.tong@damtp.cam.ac.uk}

\abstract{We describe what happens when chiral fermions scatter off magnetic monopoles. It has long been known that, in many situations, the 
outgoing state cannot be constructed from any finite number of excitations. In the language of 2d CFT, it lives in a twisted sector of the Hilbert space. Here we explain what this means from the perspective of a $d=3+1$ observer. We show that an ingoing one-particle excitation is scattered into an outgoing state surrounded by a soft cloud which consists of an infinite number of low energy particle-anti-particle pairs. We compute energy and charge correlators, showing that this cloud fragments in any measurement. A detector will register one or more hard particles which leave tracks and carry standard quantum numbers.  The remaining exotic quantum numbers live in the soft cloud, and are characterised by large charge fluctuations in the detector. }

\begin{document}
\pagestyle{plain} \setcounter{page}{1}
\newcounter{bean}
\baselineskip16pt \setcounter{section}{0}

\maketitle

\section{Introduction}

Interesting things happen when fermions scatter off monopoles  \cite{r0,r,c0,c}. In chiral theories, like the Standard Model, there is a particularly intriguing puzzle: an incoming fermion in the s-wave seemingly has no outgoing states that carry the same quantum numbers \cite{callan}. If you throw in, say, a massless right-handed electron, then what comes out?   A substantial literature has now grown out of this question \cite{joe,das,sagi,mlud,pancake,csaki,brennan,fock,csaki2,me,valya,brennanplus,me2,csaki3,cliff,lt,hook,brennan2,bpheno,ltt,frank,yuji,rishi,cop,yunquin,max,kohmori,george,craig,clay,haogeng,frank2}.

\para
The chiral nature of the problem is crucial and, as such, the puzzle only arises for massless fermions.  Give the fermions a mass, and the chiral symmetry is broken and the problem evaporates. This suggests that the answer to the puzzle lies in soft modes, carrying charge but little energy  and without a clear particle interpretation. In this spirit, Polchinski referred to the outgoing radiation as ``propagating pulses of vacuum polarization''  \cite{joe}. In part, the  purpose of this paper is to explain what those words mean.

\para
Since the scattering is in the s-wave, the problem can be studied by reducing to $d=1+1$ dimensions. Here there's little immediate sign of any soft propagating vacuum polarisation. Instead, a particle goes in and a particle comes out. The outgoing particle carries the same quantum numbers as the ingoing particle even though the perturbative Fock space doesn't allow for outgoing particles with those quantum numbers. Instead, the outgoing particle lives in a {\it twisted sector} of the Hilbert space \cite{mlud,fock,me}. More generally, the monopole acts as a non-invertible symmetry defect \cite{me,rishi,cop,yunquin,max,kohmori}. The challenge is to convert these rather formal mathematical statements about 2d conformal field theories into physical statements about what you see if you place the  monopole inside a particle detector. What really comes out?

\para
A particularly striking calculation was performed by Tachikawa, Tsuji and Watanabe \cite{yuji}. They studied a wavepacket in $d=1+1$ dimensions and showed that it emerges unscathed from the collision with the monopole/defect.  There is no dispersion, and the wavepacket exits with the same shape it entered. But, despite appearances,  there is something different about it. A wavepacket initially constructed from single particle states will emerge with support on multi-particle states. Indeed, the total number of particles and anti-particles in the outgoing wavepacket diverges.

\para
In Section \ref{scattersec}, we revisit this calculation. We compute the number of particles $n(\omega)$ and anti-particles $\bar{n}(\omega)$ in the outgoing wavepacket as a function of their energy (or frequency) $\omega$. We show that both exhibit an infra-red divergence
\be n(\omega), \bar{n}(\omega) \sim \frac{1}{\omega}\ .\label{11}\ee
This is telling us that any outgoing particle is accompanied by a soft cloud of particle-anti-particle pairs.\footnote{There is an interesting connection to the recent work \cite{david} which studies (among many other things) the {\it high energy} spectrum of particles of the Ising model in a state created by a  spin operator (which is a twist operator). The  particle spectrum in the Ising model coincides with our particle counting $n(\omega)$ for $N=4$ fermions scattering off a monopole.} This is a consequence of living in the  ``twisted sector'' and the unconventional quantum numbers of the outgoing state will reside within this cloud.

\para
In Section \ref{corrsec} we ask what this looks like in $d=3+1$ dimensions. We do so by computing energy and charge correlators introduced in \cite{diego}. We calculate the energy and charge deposited in a detector at various angles and over various time periods. 
Because the outgoing state lies in the s-wave, the angular dependence is rather trivial but, nonetheless, contains important information: it tells us that the outgoing state fragments when measured. The fact that the outgoing state is a many-particle excitation now reveals itself, with radiation going out in different directions. Some of those triggers are clearly particle-like: they carry a significant fraction of the energy, localised in some small angular region. Such radiation corresponds to an excitation in the boring, traditional Hilbert space of the theory. There is nothing exotic or twisted in the particles the detector sees.

\para
But the full charge seen by the detector coincides with that of the  incoming particle, and the scattering paradox is the statement that this charge cannot be accounted for by the usual particles alone.  There must be something more. That something can be found in soft modes, excitations with arbitrarily small energy.  These too are seen by the detector and, as we will see, are characterised by their large fluctuations. While the charge seen by the entire detector doesn't change, charge deposited in any finite region exhibits fluctuations which grow logarithmically as the observation time is increased, showing that the outgoing radiation includes correlations between very early and very late times. This is the physical manifestation of the twisted sector.

\subsection*{The Plan of the Paper}

In Section \ref{scattersec}, we start with the basics, reviewing the physics of fermions scattering off monopoles and why the s-wave is special. We cook up a situation in which a one-particle s-wave wavepacket is sent into the monopole. Standard bosonisation methods allow the construction of the outgoing s-wave, and the goal is to understand what this means. The novel calculations are in Section \ref{diracsec}, where we explain how the outgoing state has re-arranged the Dirac sea, and in Section \ref{manysec}, where, following \cite{yuji}, we count the number of particles and anti-particles in the outgoing state. 

\para
In Section \ref{corrsec}, we turn our attention to $d=3+1$ dimensions and compute energy and charge correlators. We first define these operators and then compute one-point and two-point functions. The two-point function contains information about the fragmentation of the outgoing state, since a multi-particle state can cause two disjoint regions of the detector to both register. It is here that we also see the logarithmic divergence in charge fluctuations (but not in energy fluctuations), which is the manifestation of the soft cloud of particles and anti-particles.

\para
In Section \ref{summarysec}, we attempt to summarise. We pull out the key results from the preceding sections and  explain what they  mean for the monopole-fermion scattering paradox in chiral theories.

\para
There are  three appendices. Appendix \ref{levsec} is particularly instructive: it describes a simple condensed matter system which exhibits closely analogous behaviour to that seen in monopole-fermion scattering. This consists of a quantum wire in $d=1+1$ dimensions. A localised electric field is applied to the wire and then turned off, causing two excitations, with equal and opposite charges, to propagate in opposite directions. Each of these excitations typically carries a soft cloud, with the number of electrons and holes excited above the Fermi surface given by \eqn{11}. These excitations also carry fractional charge. 

\para
Throughout the bulk of the paper, we work with a particular wavepacket of Lorentzian form. This has a number of special features. In Appendix \ref{nwesec}, we show that our main results hold for arbitrary wavepackets. We do this by counting the number of particles and anti-particles with fixed total energy and show that they too obey \eqn{11}.

\para
In Appendix \ref{masssec} we look at what happens when the fermions are given a mass. This alleviates the scattering paradox and, in principle, allows for elastic one-to-one scattering preserving all quantum numbers. In practice, that's not what happens, at least when the mass is small compared to the energy of the ingoing state. The system retains a memory of the monopole-fermion scattering paradox and, even though the soft cloud is removed, scattering into multi-particle states remains more likely. 

\section{Fermion-Monopole Scattering}\label{scattersec}

In this paper, we treat the monopole as a background field in which the fermions propagate. This means that we avoid the complications associated to photons and we ignore the rotor degree of freedom of the 't Hooft-Polyakov monopole. Instead, we have free fermions moving in the bulk and the monopole acts through the Lorentz force together, importantly, with  a boundary condition at (or near) the origin. As we will see, this simple model contains surprising subtleties.

\para
We work with left-handed massless Weyl  fermions, $\psi_i$ and $\tilde{\psi}_i$ with $i=1,\ldots,N$. The fermions $\psi_i$ have charge $+1$ under the background $U(1)$ symmetry, while $\tilde{\psi}_i$ have charge $-1$. The action is
\be S = \int d^4x\ \sum_{i=1}^N i\psi^\dagger_i\bar{\sigma}^\mu (\partial_\mu -iA_\mu) \psi_i + \sum_{i=1}^N i\tilde{\psi}_i^\dagger\bar{\sigma}^\mu (\partial_\mu +iA_\mu) \tilde{\psi}_i\ .\ee
Here $\bar{\sigma}^\mu = (\mathds{1},-\bsigma)$. The gauge potential $A_\mu$ is a background field, given by rotationally invariant Dirac monopole sitting at the origin, so 
\be \int_{{\bf S}^2} {\bf B}\cdot d{\bf S} = 2\pi g\ \ \ {\rm with}\ \  g\in \Z\ .\ee
For example, in the usual spherical polar coordinates $(r,\theta,\phi)$, we can work with
\be A_\phi = \frac{g}{2} (1-\cos\theta)\label{Aphi}\ee
which is valid in a patch away from $\theta=\pi$. Later we will restrict to minimum magnetic charge $g=1$ but, for now, we keep general magnetic charge.

\para
In the absence of a monopole,  the action has two $U(N)$ symmetries, separately rotating the $\psi$ and $\tilde{\psi}$ fermions. This doesn't survive the presence of the monopole, where we must impose boundary conditions relating the two fermions, reducing the symmetry of the problem to $SU(N)\times U(1)$ where the $U(1)$ is coupled to the gauge potential $A_\mu$. We will give details of this boundary condition shortly, but for now we mention only that the fermions will transform in the following representations:
\be \psi: \ \ {\bf N}_{+1}\ \ \ {\rm and}\ \ \ \tilde{\psi}:\ \ {\bf N}_{-1}\ .\ee
These are the symmetries that arise from $N$ doublets of fermions interacting with a 't Hooft-Polyakov monopole.

\subsection{Angular Momentum Modes}

Our goal is to understand how fermions scatter off the monopole preserving these symmetries. Scattering proceeds by decomposing the fermions in the analog of spherical harmonics which, for us, are monopole spinor harmonics \cite{yang}. For completeness, we sketch the calculation here.

\para
We start by treating $\psi$ and $\tilde{\psi}$ as classical spinor fields. The equations of motion are the usual Weyl equation
\be i\partial_t\psi = -\bsigma\cdot\bpi\,\psi \ \ \ {\rm and}\ \ \ i\partial_t\tilde{\psi} = -\bsigma\cdot\bpi\,\tilde{\psi}\label{weyleqn} \ee
where $\bpi$ is the canonical momentum
\be \bpi = -i(\nabla -iq{\bf A})\ \ee
with $q=\pm 1$. In the presence of a monopole, the system has a conserved angular momentum ${\bf J}$  which includes three terms: the orbital angular momentum ${\bf L} = \br\times \bpi$, the spin contribution $\bsigma$, and an additional anomalous term, first pointed out by Poincar\'e \cite{poincare} that arises in the presence of a monopole
\be {\bf J} = \br\times\bpi + \frac{1}{2}\bsigma - \kappa \hat{\br}\ \ \ {\rm with}\ \ \ \kappa = \frac{1}{2}|qg|\ .\ee
As is familiar in non-relativistic quantum mechanics, we can find  simultaneous eigenfunctions of ${\bf J}^2$ and $J^3$. The eigenvalues of ${\bf J}^2$ are $j(j+1)$ and those of $J^3$ are $m= -j,\ldots,+j$. The novelty is that the minimum value of $j$ is 
\be j_{\rm min} = \kappa - \frac{1}{2} = \frac{1}{2}\left(|qg| - 1\right)\ .\label{jmin}\ee
The key to the scattering paradox is to realise that the lowest angular momentum mode, with $j=j_{\rm min}$ is special. For the case of interest, with $|qg|=1$, this is an s-wave with $j_{\rm min}=0$. But it is special regardless. We will now see why.

\para
Decomposing the Weyl equations into eigenvalues of the angular momentum operator ${\bf J}^2$ allows us to turn the Weyl equation into a purely radial equation. In non-relativistic quantum mechanics, this arises because ${\bf J}^2$ coincides with the angular part of the Laplacian. Here, however, the Weyl equation is first order so we might anticipate that the angular part is related to the square-root of ${\bf J}^2$. That turns out to be more or less true. To see this, we work in the gauge \eqn{Aphi} and write the right-hand side of the Weyl equations \eqn{weyleqn} as
\be \bsigma\cdot\bpi = -i\sigma_r\partial_r - \frac{i}{r}\left[\sigma_\theta\partial_\theta + \frac{\sigma_\phi}{\sin\theta} {\cal D}_\phi\right]\ \label{bsigma} \ee
where we've defined $\sigma_r = \hat{\br} \cdot\bsigma$ and $\sigma_\theta = \hat{\btheta}\cdot\bsigma$ and $\sigma_\phi = \hat{\bphi}\cdot\bsigma$.
%
%
We want to relate the term in square brackets to ${\bf J}^2$. This is simplest to do by first introducing a new operator 
\be K = \bsigma\cdot{\bf L} +1\ee
where ${\bf L}= \br\times\bpi$ is the orbital angular momentum. We use $\sigma_r\sigma_\phi=-i\sigma_\theta$ and $\sigma_r\sigma_\theta = +i\sigma_\phi$ to show
\be \sigma_r K  = -\sigma_\theta\partial_\theta - \frac{\sigma_\phi}{\sin\theta} {\cal D}_\phi + \sigma_r\ .\ee
So we can write \eqn{bsigma} as 
\be \bsigma \cdot\bpi = -i\sigma_r\left(\partial_r + \frac{1}{r}\right) + \frac{i}{r} \sigma_r K\ \ee
and our task is to relate the operator $K$ to the square-root of ${\bf J}^2$. The commutation relations for the orbital angular momentum are $[L_i,L_j] = i\epsilon_{ijk}(L_k + \kappa\hat{\br}_k)$ and include the anomalous term. A short calculation shows that $(\bsigma\cdot{\bf L})^2 = {\bf L}^2 - \bsigma\cdot{\bf L} - \kappa \sigma_r$. This can then be used to show that
\be {\bf J}^2 = K^2+\kappa^2-\frac{1}{4}\ .\ee
The operator ${\bf J}^2$ has eigenvalue $j(j+1)$, so $K^2$ has eigenvalues $\lambda^2_j$, given by
\be \lambda_j^2 = j(j+1) +\kappa^2 + \frac{1}{4} = \left(j+\frac{1}{2}\right)^2 -\kappa^2\ .\ee
We know that $j=j_{\rm min},j_{\rm min}+1,\ldots$ where $j_{\rm min}$ is given in \eqn{jmin}. If we write
\be j = j_{\min} + n\ \ \ \ n=0,1,2,\ldots\ee
then the eigenvalues of $K^2$ are 
\be \lambda_j^2 = (\kappa+n)^2 - \kappa^2 = n(n+|qg|)\ .\ee
The eigenvalues of $K$ are then $\pm \lambda_j$. This is the result we needed. We are now in a position to see why the lowest angular mode, corresponding to $n=0$,  is special.

\subsubsection*{The Lowest Angular Momentum Mode}

The lowest angular momentum mode has $\lambda_{\rm j_{\rm min}}=0$. The eigenfunction of $K$ is a spinor $Y_0(\theta,\phi)$ that obeys
\be KY_0 = 0\ .\ee
This is also an eigenfunction of $\sigma_r$ and the index theorem tells us
\be \sigma_r Y_0 = {\rm sign}(qg)\,Y_0\ .\ee
In general, there are $2j_{\rm min} +1$ such solutions -- the usual degeneracy for angular momentum modes. We normalise these eigenspinors so that 
\be \int d^2\Omega \ Y_0^\dagger Y_0 = 1\ \ee
with $d^2\Omega$ the integral over the unit sphere ${\bf S}^2$. 

\para
The solution to the Weyl equation then takes the form
\be \psi(t,r,\theta,\phi) = \frac{\chi(t,r)}{r} Y_0(\theta,\phi) \ .\ee
The Weyl equation becomes 
\be \partial_t \chi = {\rm sign}(qg)\,\partial_r\chi\ .\ee
This is what's special about the lowest angular momentum modes: they're chiral. For $qg>0$, the modes are ingoing; for $qg<0$ they're outgoing.

\para
 For the minimal Dirac monopole, with $g=1$,  the lowest angular momentum modes of our fermions $\psi$ and $\tilde{\psi}$ are s-waves.  The $\psi$ fermions have charge $q=1$, and so are ingoing, while $\tilde{\psi}$ have $q=-1$ and so are outgoing. We write
 \be \psi_i(t,r,\theta,\phi) = \frac{\chi_i(t+r)}{r}\,Y_+(\theta,\phi)\ \ \ {\rm and}\ \ \  \tilde{\psi}_i(t,r,\theta,\phi) = \frac{\tilde{\chi}_i(t-r)}{r}\,Y_-(\theta,\phi)
 \ \ \ \label{psichi}\ee
 where, for the choice of gauge \eqn{Aphi}, the eigenspinors are
\be Y_+ = \frac{1}{\sqrt{4\pi}} \left(\begin{array}{c} \cos(\theta/2)\\ e^{i\phi} \sin(\theta/2)\end{array}\right)\ \ \ {\rm and}\ \ \ 
Y_- = \frac{1}{\sqrt{4\pi}} \left(\begin{array}{c} e^{-i\phi}\sin(\theta/2)\\  \cos(\theta/2)\end{array}\right)
\label{y+-}\ .\ee
which indeed obey $\sigma_rY_\pm = \pm Y_\pm$. 
 
 \para
 This is the essence of the scattering paradox. In the s-wave, the ingoing fermions $\psi$ transform under $SU(N) \times U(1)$ as ${\bf N}_{+1}$, while the outgoing fermions $\tilde{\psi}$ transform as ${\bf N}_{-1}$.  The challenge is to understand how ${\bf N}_{+1}$ fermions can transform into the ${\bf N}_{-1}$ fermions, while preserving both symmetries. This is the challenge we will turn to in later sections.

\subsubsection*{Higher Angular Momentum Modes}

Higher angular momentum modes are not chiral. For each $j>j_{\rm min}$, there are both ingoing and outgoing modes. To see this, first note that $\{K,\sigma_r\}=0$.  (You need to use $\partial_\theta \sigma_r = \sigma_\theta$ and $\partial_\phi\sigma_r = \sin\theta\sigma_\phi$ when showing this.) We can take the eigenspinors to obey
\be \sigma_r Y_\pm = \pm Y_\pm \ \ \ {\rm and}\ \ \ KY_\pm  =\pm  i\lambda_j Y_\mp\ .\ee
Each 4d Weyl spinor now decomposes into two modes
\be \psi(t,r,\theta,\phi)  = \frac{f(t,r)}{r} Y_+(\theta,\phi) + \frac{g(t,r)}{r} Y_-(\theta,\phi)\ee
and the Weyl equation becomes two coupled equations
\be (\partial_t-\partial_r)f = \frac{i\lambda_j}{r}g\ \ \ {\rm and}\ \ \ (\partial_t+\partial_r)g = \frac{i\lambda_j}{r}f\ .\ee
The two modes propagate in different directions, but are coupled by the angular momentum barrier $\lambda_j/r$. This means that scattering in the higher angular momentum modes is rather straightforward: incoming modes turn around and become outgoing modes. (In the quantum setting, this is accompanied by some phase shift.)  It's only the lowest angular momentum mode where the trouble lies. Nonetheless, we will have use of these higher angular momentum modes in Section \ref{corrsec} when computing energy and charge correlators.

\subsection{Sending In an S-Wave Wavepacket}

We can now set up the experiment. We construct an ingoing  wavepacket of $\psi$ in the s-wave, carrying quantum numbers ${\bf N}_{+1}$. It will bounce back  as a wavepacket in the  s-wave of $\tilde{\psi}$ which we want to understand how to interpret.

\para
The s-wave component of the $d=3+1$ dimensional fermion $\psi_i$ is the $d=1+1$ dimensional ingoing fermion  $\chi_i(t+r)$ defined in \eqn{psichi}. It has the decomposition
\be \chi_i(t+r) = \int_0^\infty\frac{d\omega}{\sqrt{2\pi}} \left[a_i(\omega) e^{-i\omega(t+r)} + b_i^\dagger(\omega) e^{i\omega (t+r)}\right]\label{chiexpansion}\ee
where $a_i$ and $b_i$ are annihilation operators for particles and anti-particles respectively, obeying $\{a_i(\omega),a_j^\dagger(\omega')\} = \{b_i(\omega),b_j^\dagger(\omega')\}=\delta_{ij} \delta(\omega-\omega')$. The vacuum $\ket{0}$ is annihilated by $a_i$ and $b_i$, so  $a_i(\omega)\ket{0} = b_i(\omega)\ket{0} = 0$ for all frequencies $\omega$. 

\para
We construct a one-particle wavepacket by sending in the particular fermion $\psi_a$ for some fixed $a=1,\ldots,N$,
\be {\ket{\rm in}} = \int_0^\infty d\omega\ f(\omega) a_a^\dagger(\omega)\ket{0}\ .\label{in}\ee
Here $f(\omega)$ is the profile of the wavepacket in frequency space. In real space, this wavepacket takes the form
\be \bra{0}\psi_i(t,r,\theta,\phi)\ket{{\rm in}} = \delta_{ia} \frac{Y_+(\theta,\phi)}{r}\,F(t+r)\ee
with $F(t+r)$ the Fourier transform
\be F(t+r) = \frac{1}{\sqrt{2\pi}}\int_0^\infty d\omega\ 
f(\omega) e^{-i\omega(t+r)}\ .\ee
We're going to take a very particular profile for the incoming wavepacket
\be f(\omega) = \sqrt{2L} e^{-L\omega}\ .\label{fomega}\ee
Here $L$ is the width of the wavepacket and the overall coefficient ensures that it is normalised, with $\int_0^\infty d\omega\,|f(\omega)|^2=1$. With this choice, the ingoing wavepacket can be written as
\be \ket{{\rm in}} = \sqrt{2L}\int_0^\infty d\omega\ e^{-LH}a_a^\dagger(\omega)\ket{0} = \sqrt{4\pi L} e^{-LH}\chi_a^\dagger(0)\ket{0} .\label{inn}\ee
In position space, the profile of the wavepacket $F(v)$, with $v=t+r$, has a Lorentzian form with a power-law tail
\be F(v) = \sqrt{\frac{L}{\pi}} \frac{1}{L + iv}\ \ \ \Longrightarrow\ \ \ |F|^2 = \frac{L}{\pi(L^2+ v^2)}\label{F}\ee
Meanwhile the phase is  ${\rm arg}\,F = -\tan^{-1}(v/L)$, and varies from $+\pi/2$ at the far left to $-\pi/2$ at the far right.  This kind of Lorentzian wavepacket has rather special features that, as we will see, make the equations much more tractable. These simplifications are closely related to a topic in quantum transport in condensed matter, known as {\it levitons}: we highlight this connection in Appendix \ref{levsec}. 

\para
 The Hamiltonian is 
\be H = \int_0^\infty d\omega\ \omega\sum_i\left[a_i^\dagger(\omega)a_i(\omega) + b_i^\dagger(\omega)b_i(\omega) \right]\ .\label{hammy}\ee
The wavepacket is not particularly well localised in energy: we have
\be \bra{{\rm in}} H\ket{{\rm in}} = \frac{1}{2L}\ \ \ {\rm and}\ \ \ \bra{{\rm in}}H^2\ket{{\rm in}} = \frac{1}{2L^2}\ .\label{hh2}\ee
Although the Lorentzian wavepacket \eqn{fomega} has nice properties, the general lessons that we extract are independent of the choice of wavepacket. This is explained in  Appendix \ref{nwesec}.

\subsection{Getting Back an S-wave Wavepacket}

The s-wave wavepacket bounces back as an s-wave. In this section, we derive the outgoing wavepacket. 

\para
For this, it's useful to bosonise, before returning to the fermionic picture. We work with Abelian bosonisation, rather than non-Abelian bosonisation. This means that we exchange the fermions for chiral bosons $\phi_i(z)$, and only the Cartan $U(1)^{N-1}\subset SU(N)$ is manifest, although the end result will be fully $SU(N)$ invariant. 

\para
 We'll switch between Euclidean and Lorentzian signatures. In Euclidean space, we work with the complex coordinate $z$, and introduce chiral bosons $\phi_i(z)$, $i=1,\ldots,N$, with 
\be \langle \phi_i(z)\phi_j(w)\rangle  = -\delta_{ij} \log(z-w)\ .\label{phiphi}\ee
A comment on Wick rotation. We take the Wick rotated time $\tau = it$ and  take $z=\tau -ix=i(t-x)$ where $x$ is an  ``unfolded'' spatial coordinate, defined by $x=r$ for $x>0$ and $x=-r$ for $x<0$. With this convention, in Euclidean space $\phi(z) = \phi(i(t+r))$ for $x<0$, which is an ingoing state, while $\phi(z) = \phi(i(t-r))$ for $x>0$ which is an outgoing state.

\para
The usual bosonisation dictionary introduces  the vertex operators
\be V_i(z) = \frac{1}{\sqrt{2\pi}} :e^{i\phi_i}:\ee
where we implicitly assume Klein factors which ensure that different fermions anti-commute. In this unfolded picture, a single scalar captures both the ingoing fermion $\chi_i(z)$ and the outgoing fermion $\tilde{\chi}_i(z)$, 
\be V_i(z)= \left\{\begin{array}{cc} \,\,\chi^\dagger_i(z)\ \ \ &x<0\\ \tilde{\chi}^\dagger_i(z)\ \ &x>0\end{array}\right.\ee
A general vertex operator is given by $V_{\blambda}(z) = \frac{1}{\sqrt{2\pi}}\! :\!e^{i\blambda\cdot \bphi(z)}\!:$\,, labeled by a vector $\blambda=(\lambda_1,\ldots,\lambda_N)$ with integer elements $\lambda_i$ which are positive to create particles and negative to create anti-particles. As we will see, in the bosonised language the scattering paradox is the statement that what comes back is a poorly quantised vertex operator with $\lambda_i$ that are not integers. Our job is to interpret this.

\para
We also have the currents
\be j_i(z) = \, :\!\chi_i^\dagger (z)\chi_i (z)\!: \,= \frac{i}{2\pi} \partial_z\phi_i(z)\ .\label{jphi}\ee
The currents are linear in the bosons, which makes the scattering simple \cite{callan}. (In the fermionic language, this can be seen in the language of boundary conformal field theory \cite{sagi,costas,ryu,philip1,li}.)  The in-current ${\bf j}_{\rm in}$ scatters into an out-current ${\bf j}_{\rm out}$ as
\be {\bf j}_{\rm out}  = R\,{\bf j}_{\rm in}\ \ \ {\rm with}\ \ \ R_{\ij} = \delta_{ij} - \frac{2}{N}\ .\ee
(The notation for the matrix $R$ isn't great: that $-2/N$ is subtracted from every element in $R$. We could alternatively write $R=\mathds{1} - \ft2N {\bf 1}{\bf 1}^T$ where ${\bf 1}^T=(1,1,\ldots,1)$.)  If we send in the  vertex operator $V_{\blambda}(z)$ for $x<0$, we get out the vertex operator 
\be V_{\blambda}(z) \longrightarrow V_{R\blambda}(z)\ .\ee
This does the job we want. Ingoing fermions have $U(1)$ charge $+1$, so the total charge of the ingoing state is $Q_{\rm in} = \sum_i\lambda_i$. Meanwhile, outgoing fermions have $U(1)$ charge $-1$, so the total charge is $Q_{\rm out} = -\sum_{i,j} R_{ij}\lambda_j$. The $R$-matrix ensures that $Q_{\rm out} = Q_{\rm in}$.  Similarly, the $SU(N)$ representation is unchanged: an $SU(N)$ Cartan generator takes the form $T= {\rm diag}(t_1,\ldots ,t_N)$ with $\sum_i t_i=0$ and for any such generator $\sum_i t_i\lambda_i = \sum_{i,j}t_i R_{ij}\lambda_j$.

\para
For our wavepacket \eqn{in}, we throw in a one-particle state built from $\chi^\dagger_a$, and the vertex operator $V_a$, so we will get out a wavepacket built from the vertex operator \cite{sagi}
\be V_{{\bm \nu}}=  :e^{i{\bm \nu}\cdot\bphi}:\ \ \ \ {\rm with}\ \  \nu_i=\left\{\begin{array}{cc} (N-2)/N&\ \ \ i=a \\ -2/N &\ \ {\rm otherwise}\end{array}\right.\label{nui}\ee
which, as promised, has non-integer coefficients. In \cite{me}, this was interpreted as a state in the twisted Hilbert space. The ingoing s-wave wavepacket \eqn{inn} is given in terms of vertex operators as
\be \ket{{\rm in}} = \sqrt{2L} e^{-LH}V_a(0)\ket{0}\ .\ee
The correctly normalised outgoing s-wave wavepacket is
\be \ket{{\rm out}} = \sqrt{2L} e^{-LH} V_{\bm\nu}(0)\ket{0}\ \label{out}.\ee
All higher angular momentum modes remain in their ground state. Our job is to understand what this means.

\subsection{Rearranging the Dirac Sea}\label{diracsec}

Many of the calculations that follow use, as an essential ingredient, the fermion propagator in the out-state
\be G_i(u,u') = \bra{{\rm out}} \tilde{\chi}_i^\dagger(u')\tilde{\chi}_i(u)\ket{{\rm out}}\ .\ee
Here we calculate it. Even before we use it in various applications, we will see that it has some peculiar properties which capture what it means to be in the ``twisted sector'' of the Hilbert space.

\para
We do the calculation in Euclidean space, where the relevant part of the out-state \eqn{out} can be written as 
\be e^{-LH}V_\nu(0)\ket{0} = V_\nu(-L)\ket{0}\ .\ee
Substituting the out-state \eqn{out} into the propagator, we're left with the computation of a four-point function of vertex operators
\be G_i(z,z') = \frac{L}{\pi}\bra{0}V^\dagger_{\nu}(L)V_i(z')V^\dagger_{i}(z) V_\nu(-L)\ket{0}\ .\ee
This is a standard calculation. The result is
%
%
%
\be G_i(z,z') = \frac{1}{2\pi}\frac{1}{z'-z} \left[\frac{(L-z)(L+z')}{(L-z')(L+z)}\right]^{\nu_i}\ .\ee
The memory of which fermion $\chi_i$ we're computing the propagator for sits in the exponent $\nu_i$. This, in turn, is given in \eqn{nui} and differs depending on whether $i=a$, the incoming fermion, or $i\neq a$.  In Euclidean space, the propagator needs a choice of branch cut -- this is the twist line. Going back to Lorentzian signature, we write $z=iu$, with $u=t-r$, and introduce the phase
\be   \frac{L-iu}{L+iu} = e^{-2i\theta(u)}\  \ \ {\rm with}\ \ \tan \theta(u) = \frac{u}{L}\ .\ee
The propagator involves the phase raised to the power $\nu_i$,  so it's also useful to introduce further notation and write
\be U_i(u) = e^{-2i\nu_i\theta(u)}\ .\label{uu}\ee
The resulting out-state propagator then becomes
\be G_i(u,u') = U_i(u)\, G_0(u,u')\, U^\star_i(u')\ \label{gnu}.\ee
This differs by phases from the vacuum propagator $G_0(u,u')$ given by
\be  G_0(u,u') = \bra{0}\tchi_i^\dagger(u')\tchi_i(u)\ket{0} 
= \frac{i}{2\pi(u-u'+i0)}\  .\ \ \ \label{g0}\ee
As we will see, one of the key features of the propagator in the out-state is that there is a phase difference between late and early times. 
Specifically, when evaluated at $u=u_{\rm early} \ra -\infty$ and $u'=u_{\rm late} \ra +\infty$, the propagator becomes

\be G_i(u_{\rm early},u_{\rm late})  \ra  e^{2\pi i \nu_i} G_0(u_{\rm early},u_{\rm late}) \ . \label{earlylate}\ee
This phase difference will be responsible for  several of the strange features of fermions scattering off monopoles. 

\para
The rather simple form of the propagator in the out-state \eqn{gnu} can be traced to the special form of the incoming wavepacket \eqn{F}. For other wavepackets, the propagator is more complicated, but the special feature \eqn{earlylate} persists.

\subsection*{The Propagator as a Projector}

We can start to get some idea for the meaning of the phase by recalling a simple interpretation of the propagator when viewed through the lens of the  Dirac sea: it acts as a  projector onto negative energy states.

\para
Let's first understand this statement, and then see what it means for our outgoing wavepacket. Consider the single-particle Hamiltonian  for a chiral fermion
\be h = i\partial_u\ee
with $u=t-r$.  Obviously it has energy eigenstates 
\be \phi_\epsilon(u) = \frac{e^{-i\epsilon u}}{\sqrt{2\pi}} \ \ \ {\rm where}\ \ h\phi_\epsilon = \epsilon \phi_\epsilon \ .\ee
In the Dirac picture, the vacuum has all states with $\epsilon<0$ occupied. The projector onto such states is
\be P(u,u') = \int_{-\infty}^0 d\epsilon\ \phi_\epsilon(u)\phi_\epsilon^\star(u') = \int_{-\infty}^0 \frac{d\epsilon}{2\pi} \ e^{-i\epsilon(u-u')}\ .\ee
We add a convergence factor $e^{\eta\epsilon}$, with $\eta>0$, and evaluate the integral
\be P(u,u') = \lim_{\eta \ra 0^+} \int_{-\infty}^0 \frac{d\epsilon}{2\pi} \ e^{-i\epsilon(u-u') + \eta \epsilon} = \frac{i}{2\pi(u-u'+i0^+)}\ .\ee
We see that this coincides with the vacuum propagator \eqn{g0}
\be P(u,u') = G_0(u,u')\ .\ee
Like all projection operators, the propagator obeys $G_0^2=G_0$, in the sense that
\be \int du' \ G_0(u,u')G_0(u',u'') = G_0(u,u'')\ .\ee
Now we can interpret the propagator in the out-state, $G_i(u,u')$, in the same way. What is this telling us about the rearrangement of the occupied vs unoccupied states in the Dirac sea?  The propagator \eqn{gnu} can be written as
\be G_i(u,u') = \int_{-\infty}^0 \frac{d\epsilon}{2\pi} \ \left[U_i(u) e^{-i\epsilon u}\right] \,\left[U_i(u') e^{-i\epsilon u'}\right]^\star\ .\ee
So the filled states of the Dirac sea are now $U_i(u) e^{-i\epsilon u}$. Because the additional phase $U_i(u)$, given in \eqn{uu}, depends on $u$, this mixes the positive and negative energy states.

\para
As a sanity check, we can see what happens in the case $\nu_i=1$. In this case, the propagator \eqn{gnu} can be written as
\be G_{\nu_i=1}(u,u') = G_0(u,u') + F(u)F(u')^\star\ee
where $F(u)$ is the wavepacket \eqn{F}
\be F(u) = \sqrt{\frac{L}{\pi}} \frac{1}{L+iu}\ee
This contains only positive energy modes
\be F(u) =  \sqrt{\frac{L}{\pi}} \int_0^\infty d\epsilon\ e^{-(L-iu) \epsilon}\ee
and so is orthogonal to the negative energy states. The interpretation of $G_{\nu_i=1}(u,u')$ is that, in addition to the filled negative energy states, the one-particle wavepacket $F(u)$ is also occupied. In other words, we have an excited particle.

\para
For $\nu_i=-1$, there is a similar story. The propagator is now
\be G_{\nu_i=-1}(u,u') = G_0(u,u') - F(u)^\star F(u')\ee
This is a projection onto the filled Dirac sea, but with one negative energy state removed: it is the excitation of a hole. 

\para
For fractional $\nu_i$, the propagator \eqn{gnu} can be written as
\be G_i(u,u') = G_0(u,u') + \frac{\sin(\nu_i[\theta(u)-\theta(u')])}{\pi(u-u')} e^{-i\nu_i[\theta(u) - \theta(u')]}\ .\label{dforlater}\ee
It no longer has the interpretation as the addition or removal of a single excitation. Instead, the positive and negative energy states have been mixed and the Dirac sea contains coherent superpositions of both.  This is how the ``pulse of vacuum polarisation'' \cite{joe} or the ``twisted sector'' \cite{me} shows up in this analysis. 

\para
This correlation between filled and empty states persists in the limit $u_{\rm early} \ra -\infty$ and $u_{\rm late} \ra +\infty$. From \eqn{dforlater}, we have
\be \left|G_{i}(u_{\rm early},u_{\rm late}) - G_0(u_{\rm early},u_{\rm late})\right|^2 \ra \frac{\sin^2(\pi \nu_i)}{\pi^2 (u_{\rm early}- u_{\rm late})^2}\ .\label{earlylate2}\ee
These late time correlations will ultimately give rise to infra-red divergences in various quantities. We will see the first of these when we count the number of particles excited in the outgoing wavepacket. 

\subsection{One Goes In,  Many Come Out}\label{manysec}

We sent in a wavepacket \eqn{in} built from one-particle states. But what comes out is a wavepacket without definite particle number. This is one of the consequences of living in the twisted Hilbert space: it's a state that can't be built directly from the creation operators $a_i^\dagger$ and $b_i^\dagger$. This was made more precise in \cite{yuji} where they calculated the  particle occupation numbers of a wavepacket (for the case of $N=4$). 
Here we provide an alternative version of this calculation.

\para
To start, we work with just a single fermion, and return to the full $N$ fermions later. The out-state is labelled by $\nu\in \R$ 
\be \ket{{\rm out}} = (2L)^{\nu^2/2} \,e^{-LH}V_\nu(0)\ket{0}\ \ {\rm where}\ \ V_\nu = :e^{i\nu\phi}:\ee
(The normalisation differs from \eqn{out} because we have just a single fermion while \eqn{out} describes all $N$ fermions: the normalisation in \eqn{out} could be written as $(2L)^{{\bm \nu}^2/2}$, but the vector ${\bm \nu}$ has norm ${\bm \nu}^2 = 1$.)
We want to calculate the number of excited particles $n_\nu(\omega)$ of frequency $\omega$ and, separately, the number of excited anti-particles $\bar{n}_\nu(\omega)$. These are given by
\be n_\nu(\omega) &=& \bra{{\rm out}} a^\dagger(\omega)a(\omega)\ket{{\rm out}} \nn\\\ \bar{n}_\nu(\omega) &=& \bra{{\rm out}} b^\dagger(\omega)b(\omega)\ket{{\rm out}}\ .\ee
Let's first compute the number of particles. Using the expansion \eqn{chiexpansion}, we have, for $\omega>0$,
\be a(\omega) = \frac{1}{\sqrt{2\pi}} \int_{-\infty}^{+\infty} du\ e^{+i\omega u} \tilde{\chi}(u)\  \ \ \ {\rm for}\ \ \omega>0.\ee
Here the integral is over the lightcone coordinate $u=t-r$. This gives us an expression for the particle number
\be n_\nu(\omega) = \frac{1}{2\pi}\int_{-\infty}^{+\infty} du \int_{-\infty}^{+\infty} du' \ e^{i\omega(u-u')} G_\nu(u,u')\label{nG}\ee
where $G_\nu(u,u')$ is the fermion propagator in the out-state, which we computed  in \eqn{gnu}
\be G_\nu(u,u') = U_\nu(u)\, G_0(u,u')\, U^\star_\nu(u')\ee
with 
\be U_\nu(u) = e^{-2i\nu \theta(u)} \ \ {\rm and}\ \  \tan \theta(u) = \frac{u}{L}\ .\ee
Substituting this expression into \eqn{nG}, the number of excited particles is
\be n_\nu(\omega) &=& \frac{1}{2\pi} \int du\,du'\ e^{i\omega(u-u')} e^{-2i\nu[\theta(u)-\theta(u')]}G_0(u,u')\nn\\ 
&=& \frac{1}{(2\pi)^2}\int du\,du'\int_0^\infty d\omega'\ e^{i(\omega+\omega')(u-u')} e^{-2i\nu[\theta(u)-\theta(u')]}\nn\ee
The $\int du$ and $\int du'$ integrals are both Fourier transforms of the phase $U_\nu(u)$,
\be \widehat{U}_\nu (k) = \int_{-\infty}^{+\infty} du\ e^{ikv} \left(\frac{L-iu}{L+iu}\right)^\nu\ .\ee
This gives   the expression for the particle number
\be n_\nu(\omega) = \int_\omega^\infty \frac{dk}{(2\pi)^2} |\widehat{U}_\nu(k)|^2\ .\label{nomega}\ee
A similar calculation counting anti-particles gives the result
\be \bar{n}_\nu(\omega) = n_{-\nu}(\omega)\ .\label{nnbar}\ee
To orient ourselves, it's useful to repeat the exercise from the previous section and compute these particle numbers when $\nu=\pm1$, in which case vertex operators are correctly normalised, $V_{\pm 1} =\, :\!e^{\pm i\phi}\!:$\,, and create a particle and anti-particle respectively.  For $\nu=1$, we have
\be \frac{L-iu}{L+iu} = -1 + \frac{2L}{L+iu}\ .\ee
Its Fourier transform is
\be \widehat{U}_1(k) =  -\int_{-\infty}^{+\infty} du\ e^{iku}  + 2L\int_{-\infty}^{+\infty} du\ \frac{e^{iku}}{L+iu} = -2\pi\delta(k) + 4\pi L e^{-Lk} \Theta(k)\ \ \ \ee
with $\Theta(k)$ the step function. The integral in \eqn{nomega} picks up contributions only for $k\geq \omega>0$, so the delta function doesn't contribute, and we get
\be n_1(\omega) = 2Le^{-2L\omega}\ .\ee
This is the particle density of the incoming wavepacket.  Indeed, the  total particle number is 
\be N_{\rm particle} = \int_0^\infty d\omega \ n_1(\omega) = 1\ .\ee
The energy of this wavepacket is $\langle H\rangle = \int d\omega \,\omega n_1(\omega) = 1/2L$, in agreement with \eqn{hh2}.

\begin{figure}[htb]
\begin{center}
\epsfxsize=6in\leavevmode\epsfbox{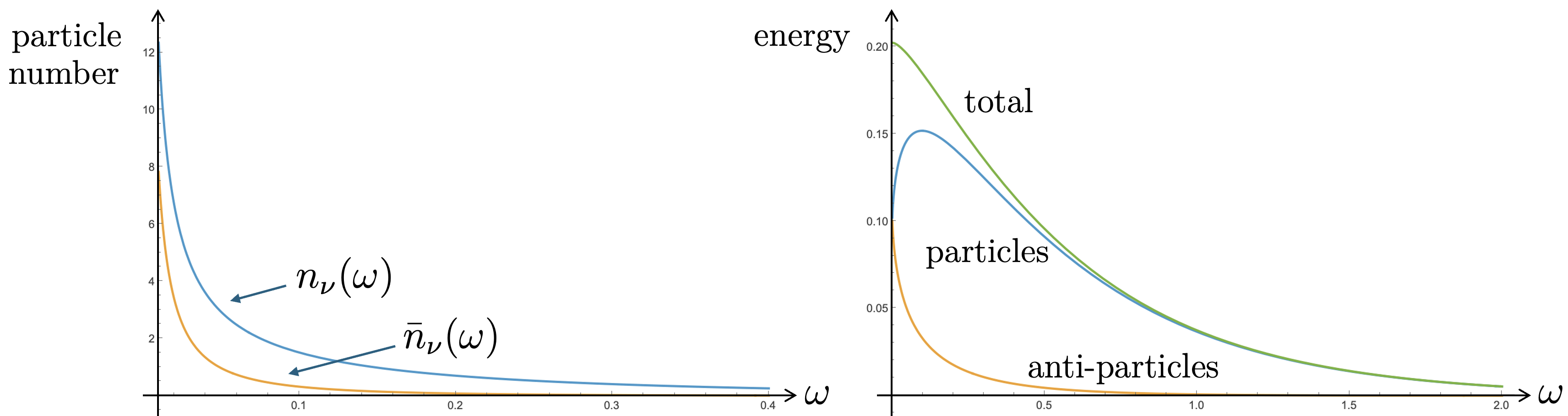}
\end{center}
\caption{The particle spectrum (on the left) and energy spectrum (on the right) for $\nu=1/2$ (which is relevant for the case $N=4$). Other values of $\nu\in (0,1)$ have similar properties.}
\label{nnbarfig}
\end{figure}

\para
The same calculation for $\nu=-1$ gives
\be \widehat{U}_{-1}(k) = -2\pi \delta(k) + 4\pi Le^{Lk}\Theta(-k)\ .\ee
In this case, the function has non-trivial support only for $k\leq 0$, but this doesn't contribute to the integral \eqn{nomega}, meaning that we have $n_{-1}(\omega)=0$. This reflects the fact that the vertex operator $V_{-1}$ creates an anti-particle, instead of a particle. Indeed, if we look at the anti-particle excitations, using \eqn{nnbar}, we have $\bar{n}_{-1}(\omega) = n_1(\omega) = 2L^{-2L\omega}$, while $\bar{n}_1(\omega)=0$.

\para
More interesting is what happens when $\nu$ is not an integer. Now both the particle and anti-particle spectrum are non-zero. Their form is shown on the left of  Figure \ref{nnbarfig}. They are exponentially suppressed at large $\omega$, but diverge at small $\omega$,
\be n_\nu(\omega) \sim \bar{n}_\nu(\omega) \sim \frac{\sin^2(\pi\nu)}{\pi^2\omega}\ \ {\rm as}\ \ \omega\ra 0\ .\label{nnbardiverge}\ee
This means that the total number of both particles and anti-particles has an infra-red log divergence, as seen in \cite{yuji}. It is telling us that an outgoing state with non-integer $\nu$ has a cloud of particle-hole pairs. There is no such divergence in the total energy: it is given by $\langle H\rangle = \nu^2/2L$. The energy of particles, anti-particles, and their sum, as a function of $\omega$, is shown on the right of Figure \ref{nnbarfig}.

\subsection*{Charge Accounting}

The discussion above was for a single particle. For our set-up, we throw in a fermion $\chi_a(u)$ and get out fermions $\tilde{\chi}_i$ with particle density $n_{\nu_i}(\omega)$, where $\nu_i$ is given in \eqn{nui}.

\para
Here there is a puzzle. Each of the outgoing fermions $\tilde{\chi}_i$ carries charge $-1$, while the anti-particles carry charge $+1$. That means that the $i^{\rm th}$ species of fermion carries charge
\be q_i(\omega) = -n_{\nu_i}(\omega) + \bar{n}_{\nu_i}(\omega)\ .\ee
The leading divergence \eqn{nnbardiverge} is the same for particles and anti-particles, but there is a sub-leading term in the low-frequency expansion which gives
\be q_i(\omega) \sim \frac{4L}{\pi^2}\sin^2\left(\frac{2\pi}{N}\right) \log^2\left(\frac{1}{L\omega}\right) + \ldots\ .\ee
This is an integrable divergence. It is telling us that the soft cloud, comprising of infinite number of particle-anti-particle pairs, can carry finite residual charge. 

\para
We can also calculate the total charge carried by outgoing particles and anti-particles of all frequencies
\be Q_{\rm particles} =   \sum_{i=1}^N \int_0^\infty d\omega\ q_i(\omega) \ .\ee
A short calculation gives
\be Q_{\rm particles}  = \frac{i}{2\pi} \int du\ \sum_{i=1}^N U^\star_{\nu_i}(u) U'_{\nu_i}(u) 
 + \frac{1}{2\pi} \sum_{i=1}^N {\rm Im}\left[U_{\nu_i}(-\infty) U^\star_{\nu_i}(+\infty)\right]\label{imuu} \ee
from which it's easy to calculate
\be Q_{\rm particles}  = -\sum_{i=1}^N\left(\nu_i - \frac{\sin(2\pi\nu_i)}{2\pi}\right)  = 1-\frac{N}{2\pi}\sin\left(\frac{4\pi}{N}\right)\ .\ \ \ \label{missingcharge}\ee
That's odd. The ingoing charge was $+1$, but there seems to be a shortfall in the outgoing particles. This shortfall vanishes for $N=4$, but increases with $N$. As $N\ra \infty$, the total charge carried in these particles is $Q_{\rm particles} = -1$, a deficit of 2. 

\para
We can also look at the flavour charges, labelled by Cartan elements $T= {\rm diag}(t_1,\ldots,t_N)$ with $\sum_it_i=0$. Here we have
\be Q({\bf t})_{\rm flavour} = -\sum_{i=1}^N \int_0^{\infty} d\omega\ t_iq_i(\omega) = t_a\ee
So there's no discrepancy in the outgoing flavour charges: only in the overall charge $Q_{\rm total}$. What's going on?

\para
The quantities $n(\omega)$ and $\bar{n}(\omega)$ count the number of particle and anti-particle excitations  above the usual ground state, but these aren't sufficient to count for the full charge of the outgoing state. Yet we constructed this outgoing state so that it carries the same charge, $Q=+1$ as the ingoing state, a fact that we will confirm in the next section when we measure the outgoing charge in a different way. The calculation above is telling us that this charge is not solely carried by particle-like excitations. Instead, some fraction of it sits in a something more nebulous.

\para
Technically, this deficit can be traced to a kind of boundary term -- the second term in  \eqn{imuu} -- which is non-vanishing because of the phase difference between $U_\nu(-\infty)$ and $U_\nu(+\infty)$. A different way to view this is that we have constructed the calculation above to count excitations with frequency $\omega >0$. The phase difference between very early and late times is telling us that the missing charge can be found as a delta-function contribution at $\omega=0$.

\para
In the next section, we'll revisit these issue through the lens of energy and charge correlators.   We'll further see how the early-late phase correlation that is responsible for the missing charge also  manifests itself in growing long-time fluctuations of the measurement of charge.  This, ultimately, is the key to  the original scattering paradox. Most of the outgoing energy is carried by particles, while the charge is more diffuse.

%
%
%

\section{Charge and Energy Correlators}\label{corrsec}

To further make sense of fermion-monopole scattering, we turn to energy and charge correlators. These are related to (highly idealised!) detectors: we measure the amount of charge and energy deposited in regions in a sphere surrounding the monopole. 

\para
In this section, we first introduce the observables. In Section \ref{onepointsec}, we compute the one-point correlation functions, These are rather trivial (although they do help resolve the missing charge seen in the previous section). In Section \ref{twopointsec}, we compute two-point correlation functions. This is where the fact that we live in $d=3+1$ dimensions, rather than $d=1+1$ dimensions, becomes important, because even though the outgoing state is in the s-wave, higher angular momentum modes play a role as intermediate states. This means that we can see the outgoing fermion scatters in multiple directions, as a soft shower, rather than a single hard outgoing particle.

\para
There is a current associated to each of the $i=1,\ldots,N$ fermions in   $d=3+1$ dimensions
\be J^\mu_i =\,  :\!\psi_i^\dagger \bar{\sigma}^\mu \psi_i:  - :\!\tilde{\psi}_i^\dagger \bar{\sigma}^\mu \tilde{\psi}_i:\ .\ee
The fermions $\psi_i$ are in their vacuum state and won't contribute to the connected correlators of interest.  The fermions $\tilde{\psi}_i$ have an expansion in monopole harmonics. For the outgoing modes, this reads
\be \tilde{\psi}_i = \frac{1}{r}\left[\tilde{\chi}_i(u) Y_0(\theta,\phi) + \sum_{l=1}^\infty \tilde{\chi}_{i,l}(u) Y_l(\theta,\phi)\right]\ee
with $u=t-r$. Here $Y_0$ is the s-wave harmonic: it is given by  $Y_0=Y_-$ in \eqn{y+-}. The higher angular momentum modes include an implicit $m=-l,\ldots,+l$ azimuthal quantum number that we have suppressed. In what follows, we will also use the notation $\tilde{\chi}_{i,0} = \tilde{\chi}_i$ for the s-wave fermions. 

\para
We're interested in the radial current $J_i=J^r_i$.  Each of the  monopole harmonics obeys $\sigma_rY_l=-Y_l$ with  $l=0,1,2,\ldots$ for outgoing fermions. The upshot is that the radial current is given by
\be J_i (t,r,\theta,\phi) = -\frac{1}{r^2} \sum_{l,l'} Y_l^\dagger(\theta,\phi) Y_{l'}(\theta,\phi) \,:\!\tilde{\chi}^\dagger_{i,l}(u) \tilde{\chi}_{i,l'}(u):\ .\ee
The s-wave is excited in the state $\ket{{\rm out}}$ than we constructed in the previous section. Meanwhile, all of the higher angular momentum modes $\tilde{\chi}_{i,l}$ with $l=1,2,\ldots$ sit in their ground state. Nonetheless, they will be important when we compute two-point functions, where they appear as intermediate states.

\begin{wrapfigure}{r}{0.3\textwidth}
\centering
\includegraphics[width=.7\linewidth]{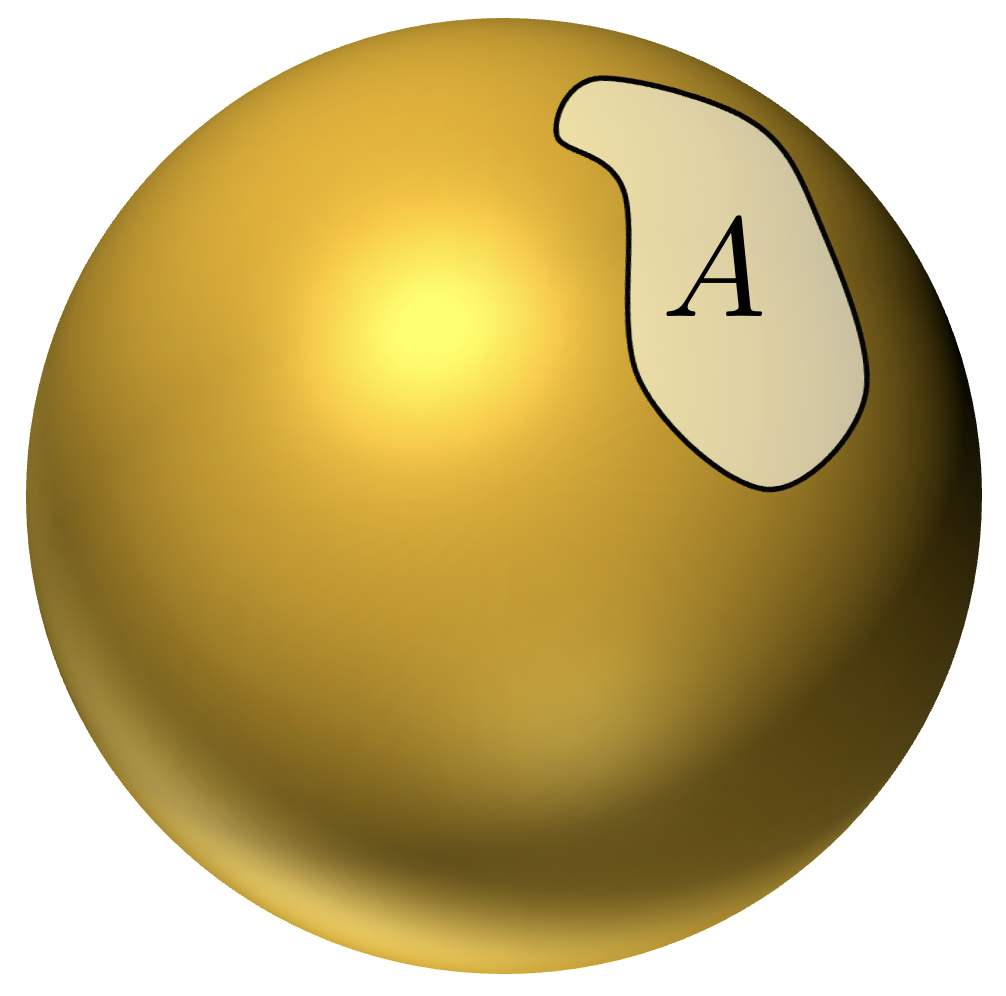}
\end{wrapfigure}

\para
We're interested in the current through a region $A$ on the sphere that surrounds the monopole
\be I_{i,A}(u) = r^2 \int_A d^2\Omega\ J_i(t,r,\theta,\phi)\ .\ee
The total $U(1)$ current is
\be I_A(u) = \sum_{i=1}^N I_{i,A}(u)\ .\ee
Meanwhile, the flavour currents, associated to $SU(N)$, are given by $\sum_i t_i I_{i,A}$ with $\sum_i t_i=0$.

\para
We will also be interested in the energy deposited in the region $A$. The appropriate component of the stress tensor is
\be T^{0r}(t,r,\theta,\phi)  = \frac{i}{2r^2} \sum_{i=1}^N \sum_{l,l'} Y_l^\dagger(\theta,\phi)Y_{l'}(\theta,\phi) \,:\left[\tilde{\chi}_{i,l}(u)\partial_u
\tilde{\chi}_{i,l'}(u) - \partial_u\tilde{\chi}_{i,l}\,\tilde{\chi}_{i,l'}(u)\right]:\ .\ee
The energy deposited in the region $A$ is
\be {\cal E}_A(u)  = r^2 \int_A d^2\Omega\ T^{0r}(r,t,\theta,\phi)\ .\label{calE}\ee
Our goal is to compute correlation functions of $I_{i,A}$ and ${\cal E}_A$ and interpret what these are telling us about monopole-fermion scattering.

\subsection{One-Point Correlation Functions}\label{onepointsec}

We start with computing the one-point functions. These are somewhat trivial, although they do highlight an issue seen in the missing charge in the particle count of $n(\omega)$ and $\bar{n}(\omega)$ of Section \ref{manysec}. 

\para
We start with the current associated to the fermion $i$ through a region $A$. This is
\be I_{i,A}(u) =  \int_A d^2\Omega\  J_i (t,r,\theta,\phi) = - \sum_{l,l'} Y_l^\dagger(\theta,\phi) Y_{l'}(\theta,\phi) \,:\!\tilde{\chi}^\dagger_{i,l}(u) \tilde{\chi}_{i,l'}(u)\!:\ .\label{i}\ee
We want to compute the expectation value of this current in the excited s-wave state that arises from throwing in a fermion of species $i=a$
\be \ket{{\rm out}} = \sqrt{2L} \,e^{-LH}V_{{\bm \nu}}(0)\ket{0}\ \ {\rm where}\ \ \nu_i = \delta_{ia} - \frac{2}{N}\ .\ee
All higher angular momentum modes $\tilde{\chi}_{i,l}$ with $l=1,2,\ldots$ sit in their ground state which means that they don't contribute to the current. We then have
\be \langle I_{i,A}(u)\rangle_{\rm out} = -\int_A d^2\Omega\ Y_0^\dagger(\theta,\phi) Y_0(\theta,\phi)\,\langle :\!\tilde{\chi}^\dagger_{i}(u) \tilde{\chi}_{i}(u)\!:\rangle_{\rm out}\ .\ee
It will be useful for what follows to define the normal ordered propagator 
\be D_i(u,u') = \langle :\!\tilde{\chi}^\dagger_{i}(u') \tilde{\chi}_{i}(u)\!:\rangle_{\rm out}\ .\ee
We don't have to do any more work to compute this. The normal ordering is 
\be :\!\tilde{\chi}^\dagger_{i}(u') \tilde{\chi}_{i}(u)\!:\ = \tilde{\chi}^\dagger_{i}(u') \tilde{\chi}_{i}(u) - \bra{0} \tilde{\chi}^\dagger_{i}(u') \tilde{\chi}_{i}(u)\ket{0}  = \tilde{\chi}^\dagger_{i}(u') \tilde{\chi}_{i}(u) -  G_0(u,u')\ee
so we have
\be D_i(u,u') = G_i(u,u') - G_0(u,u') = \frac{\sin(\nu_i[\theta(u)-\theta(u')])}{\pi(u-u')} e^{-i\nu_i[\theta(u) - \theta(u')]}\label{dj}\ee
where we've used our previous result \eqn{dforlater}. Now we've got everything we need. As $u'\ra u$, we have $\theta(u) -\theta(u') = (u-u')\theta'(u) + \ldots$, and so
\be D_i(u,u) = \frac{\nu_i}{\pi} \frac{L}{L^2+u^2}\ .\ee
This is the profile of the wavepacket $|F(u)|^2$ that we introduced in \eqn{F}
\be D_i(u,u) = \nu_i |F(u)|^2\ \ \ {\rm with}\ \ |F(u)|^2 = \frac{L}{\pi(L^2+ u^2)}\label{Fagain}\ee
That makes sense:  the outgoing current takes the same shape as the incoming wavepacket. Integrating over the region $A$ of the sphere ${\bf S}^2$ gives a trivial factor of $|A|/4\pi$, where $|A|$ is  the solid angle of $A$. (So, for example,  $|A|=4\pi$ for the whole sphere, or $|A|=2\pi$ for the hemisphere.) The end result is
\be \langle I_{i,A}(u)\rangle_{\rm out} =- \frac{|A|}{4\pi} \nu_i |F(u)|^2\ .\ee
The total current,  $I_A = \sum_i I_{i,A}$, is given by
\be
\langle I_{A}(u)\rangle_{\rm out} = \frac{|A|}{4\pi}|F(u)|^2\ .\ee
Suppose that we run the measurement for a time $-T<u<T$. Then  the charge that passes through the region $A$ in that time is
\be \langle Q_A(T)\rangle_{\rm out} = \frac{|A|}{4\pi} \int_{-T}^{+T} du\ \frac{L}{\pi(L^2+u^2)} = \frac{|A|}{2\pi^2} \tan^{-1}\left(\frac{T}{L}\right) \ .\label{qat}\ee
Wait long enough, so $T\ra \infty$, and we measure charge $Q_A\ra |A|/4\pi$. In particular, the total charge that passes through the sphere $A={\bf S}^2$ is simply $Q= 1$. In many ways that's unsurprising: it's the same as the charge that we threw in. It's noticeable only because it differs from the charge we found in \eqn{missingcharge} where we counted particle excitations.

%
%
%
%
%
%
%
%
%
%
%
%

\subsection*{The Energy One-Point Function}

The one-point function of the energy observable can be computed straightforwardly in the same manner. We are interested in the energy deposited in a region $A$
\be {\cal E}_A(u) = \frac{i}{2} \sum_{i,l,l'} \int_A d^2\Omega\ Y_l^\dagger(\theta,\phi)Y_{l'}(\theta,\phi) :\left[\tchi^\dagger_{i,l}(u)\partial_u \tchi_{i,l'}(u) -  \partial_u\tchi_{il}^\dagger\,\tchi_{i,l'}(u)\right]:\ .\ee
It differs from the current by those derivatives. The one-point function is 
\be \langle {\cal E}_A(u)\rangle = \frac{|A|}{4\pi} \sum_i \frac{i}{2} \lim_{u\ra u'} (\partial_u-\partial_{u'})D_i(u,u') =  \frac{|A|}{4} |F(u)|^4\ee
with $|F(u)|^2$ the profile of the wavepacket \eqn{Fagain}. Picking $A={\bf S}^2$, and integrating over all time gives the total energy deposited on the sphere as
\be \langle E\rangle = \int_{-\infty}^{+\infty}du\ \langle {\cal E}_{{\bf S}^2}(u)\rangle  = \frac{1}{2L}\ .\ee
This coincides with the expectation value of the Hamiltonian evaluated on the incoming wavepacket \eqn{hh2}.

%
%
%
%

\subsection{Two-Point Functions and Fragmentation}\label{twopointsec}

We can extract further information from the two-point functions of charge and energy correlators. Typically, these  correlators capture the angular dependence of complicated showers of outgoing particles. But, for us, that angular dependence is trivial, because the outgoing state is in the s-wave. Nonetheless, there is information lurking in the two-point function. Before we calculate it, we explain what we're looking for.

\para
To do so, let's first step back from our problem of interest and consider a simpler situation where $M$ massless charged scalar particles emerge from a scattering event. The individual particles carry charges $q_i$, which can be positive or negative, while the total charge is fixed
\be \sum_{i=1}^M q_i = Q\ .\ee
There's a probability distribution for the particles to go in different directions. We'll write this as $P(\Omega_1,\ldots,\Omega_M)$, where $\Omega_i=(\theta_i,\phi_i)$ labels the direction of the $i^{\rm th}$ particle.

\para
We will measure the expectation values of charge operators, which tell us how much charge is detected in a region $A$  on the sphere. In this simplified setting, we can write these as 
\be Q_A = \sum_{i=1}^M q_i\,\mathbf{1}_A(\Omega_i)\ .\ee
Here $\mathbf{1}_A(\Omega)= 1$ when $\Omega\in A$ and is zero otherwise. If we take $A$ to be the whole sphere, we get $Q_{A = {\bf S}^2}=Q$, the total charge.  The two point-function of these charge observables is then
\be
\langle Q_A Q_B\rangle = \int \prod_{i=1}^M d^2\Omega_i\ P(\Omega_1,\ldots,\Omega_M) \sum_{j,k}q_jq_k\, \mathbf{1}_A(\Omega_j)\,\mathbf{1}_B(\Omega_k)\ .
\ee
Now restrict to the situation where each particle emerges in the s-wave. In this case, the probability distribution is uniform over the sphere and we have
\be
P(\Omega_1,\ldots,\Omega_M) = \frac{1}{(4\pi)^M}\ee
where the factors of $1/4\pi$ ensure that this is normalised.  The two-point function now simplifies dramatically. There are two different kinds of contributions. When $j=k$, the same particle contributes to both measurements, which only happens if there is a non-vanishing  overlap region $A\cap B$. When $j\neq k$,  two particles can go in different directions and still contribute. The upshot is that 
\be
\langle Q_A Q_B\rangle =  \frac{|A\cap B|}{4\pi} \sum_i q_i^2 + \frac{|A|\,|B|}{(4\pi)^2} \sum_{i\neq j}q_iq_j 
\ee
We see that for particles in the s-wave --- which is the situation of interest in this paper --- the charge correlator contains just two terms. Moreover, the coefficients of the two terms are related by charge conservation
\be
Q^2=\sum_iq_i^2+\sum_{i\neq j}q_iq_j\ .
\ee
We introduce the number
\be
\gamma = Q^2 -  \sum_i q_i^2 = \sum_{i\neq j} q_iq_j \ .\ee
Then the two point function takes the form
\be
\langle Q_A Q_B\rangle =  (Q^2-\gamma)  \frac{|A\cap B|}{4\pi}  + \gamma \frac{|A|\,|B|}{(4\pi)^2} \ . \label{gammalesson}\ee
The only information in this correlation function lies in the single number $\gamma$.  If the charge is emitted as a single particle, so  $M=1$,  we have $\gamma=0$. If, instead, there are $M$ outgoing particles each of charge $q_i=1$, then $\gamma = M(M-1)$. And if there are $M$ particles with charges $q_i=\pm 1$ so that the overall state is neutral, then $\gamma = -M$. In this way, the number $\gamma$ is a crude way to capture the fragmentation of the outgoing state into different particles, each with the ability to move in different directions. This is what we will compute for our fermion-monopole scattering problem, together with analogous expressions for energy correlators.

\subsection{Current Correlators}

We start with the two-point function of current correlators (which are somewhat easier than energy correlators). Correlations of the total current (as opposed to individual flavour currents) are given by
\be \langle I_A(u) I_B(u')\rangle_{\rm out} &=& \sum_{j,l,m}\sum_{j',l',m'} \left[\int_A d^2\Omega\ Y_l^\dagger Y_m\right]\,\left[\int_B d^2\Omega'\ Y^\dagger_{l'}Y_{m'}\right] \nn\\ 
&&\ \ \ \ \ \ \ \ \ \ \ \ \times\ \langle :\!\tchi^\dagger_{j,l}(u)\tchi_{j,m}(u)\!:\, :\!\tchi^\dagger_{j',l'}(u')\tchi_{j',m'}(u')\!:\rangle_{\rm out}
\ee
The 4-point function can be computed using Wick contractions. (Usually Wick contractions are only valid in the vacuum state; they hold in this case because the propagator in the state $\ket{\rm out}$  differs from that in vacuum only by a phase \eqn{gnu}.) This gives
\be \langle :\!\tchi^\dagger_{j,l}(u)\tchi_{j,m}(u)\!:\, :\!\tchi^\dagger_{j',l'}(u')\tchi_{j',m'}(u')\!:\rangle_{\rm out} &=& \langle :\!\tchi^\dagger_{j,l}(u)\tchi_{j,m}(u)\!:\rangle_{\rm out}\,\langle :\!\tchi^\dagger_{j',l'}(u')\tchi_{j',m'}(u')\!:\rangle_{\rm out}\nn\\
&&\ +\ \langle \tchi^\dagger_{j,l}(u)\tchi_{j',m'}(u')\rangle_{\rm out}\,\langle\tchi_{j,m}(u)\tchi^\dagger_{j',l'}(u')\rangle_{\rm out}\ .
\nn\ee
The first, disconnected, term is normal ordered while the second, connected, term is not. The disconnected piece receives contributions only from the s-wave and simply gives $\langle I_A(u)\rangle_{\rm out}\langle I_B(u')\rangle_{\rm out}$. Meanwhile, each connected piece comes with two terms
\be \langle \tchi_{j,l}^\dagger(u) \tchi_{j',m'}(u')\rangle_{\rm out} = \delta_{jj'}\delta_{lm'} G_0(u',u) + \delta_{jj'} \delta_{l0}\delta_{m'0} D_j(u,u')\ .\ee
The first term is telling us that each angular momentum mode propagates normally; the second term captures the correction to the s-wave propagation because we're in the state $\ket{\rm out}$. We have two such connected propagators multiplied together, meaning that we get four terms in total. (In fact, there is a fifth term because we get an extra $\delta(u-u')$ from reordering the $\tchi(u)$ and $\tchi^\dagger(u')$ in the second connected propagator. That cancels another contact term that we will describe below.) The term proportional to $G_0^2$ gives a contribution that captures the charge fluctuations in the vacuum $\langle I_A(u)I_B(u')\rangle_0$. That leaves the $D_j^2$ term and the cross-terms $G_0D_j$.  

\para
The term proportional to $D_j^2$ has straightforward angular dependence and gives a contribution to the current-current correlator of the form 
\be -\frac{|A|\,|B|}{(4\pi)^2} \sum_j D_j(u',u) D_j(u,u')\ .\ee
The cross-terms have a more interesting structure. They take the form
\be \mbox{cross-terms} = -\sum_j \sum_l &&\!\left(\left[\int_A d^2\Omega\ Y_0^\dagger Y_l\right]\,\left[\int_B d^2\Omega'\ Y_l^\dagger Y_0\right] G_0(u,u')D_j(u',u) \right.\nn\\ &&\ \ \ \ \left. +\  \left[\int_A d^2\Omega \ Y_l^\dagger Y_0\right]\,\left[\int_B d^2\Omega'\ Y_0^\dagger Y_l\right] G_0(u',u)D_j(u,u')  \right) \nn\ee
But the spherical harmonics $Y_l(\theta,\phi)$ are a complete basis of functions on the sphere. (Recall that the $l$ index here implicitly includes the azimuthal quantum number.) If we denote $\Omega=(\theta,\phi)$, then we have the completeness relation (within the appropriate spinor subspace) 
\be \sum_l Y_l(\Omega) Y_l^\dagger(\Omega')  = \delta^2(\Omega - \Omega')\ .\ee
This means that the angular integration in the cross-terms is only non-vanishing when the regions of $A$ and $B$ overlap, and we get
\be \mbox{cross-terms} = -\sum_j\frac{|A\cap B|}{4\pi} \left[G_0(u,u')D_j(u',u) + G_0(u',u)D_j(u,u')\right]\ee
Finally, a quick calculation shows that 
\be G_0(u,u')D_j(u',u) + G_0(u',u)D_j(u,u') = \delta(u-u') D_j(u,u) -D_j(u,u')D_j(u',u) \ .\nn\ee
That delta function cancels a similar contact term described above.  Putting it all together, we get the final result for the charge fluctuations
\be  \langle I_A(u) I_B(u')\rangle_{\rm out} &=&  \langle I_A(u) I_B(u')\rangle_0 + \langle I_A(u)\rangle_{\rm out} \langle I_B(u')\rangle_{\rm out}\nn\\ &&\ \ +\ \left[\frac{|A\cap B|}{4\pi} - \frac{|A|\,|B|}{(4\pi)^2}\right]\sum_j |D_j(u,u')|^2\ .\label{iaib}\ee
%
%
%
We can use the expression \eqn{dj} for $D_j(u,u')$ to write this as
%
%
%
\be  \langle I_A(u) I_B(u')\rangle_{\rm out}  &=& \langle I_A(u) I_B(u')\rangle_0 \label{ii}\\ &&+  \  \left[ (1-\gamma(u,u')) \frac{|A\cap B|}{4\pi} + \gamma(u,u') \frac{|A|\,|B|}{(4\pi)^2} \right] |F(u)|^2|F(u')|^2 \nn\ee
where 
\be \gamma (u,u')= 1- \sum_{j=1}^N \left(\frac{\sin(\nu_j[\theta(u') - \theta(u)])}{\sin(\theta(u) - \theta(u'))}\right)^2 \ .\ee
This is our final result for the fluctuations, with the function $\gamma(u,u')$ analogous to the number $\gamma$ in \eqn{gammalesson} that captures the fragmentation of the outgoing state.  Let's now think about what it means. 

\para
First, consider the cases $N=1$ and $N=2$. Here there is no scattering paradox and in both cases we have $\gamma=0$ which is telling us that the correlations are proportional to $|A\cap B|$ and hence vanish between disjoint regions. This reflects the fact that the outgoing state is a single particle. But for general $N$, where we have a scattering paradox, the situation is more interesting.

\para
First note that if we set $u=u'$ then we have $\gamma(u,u)=0$. This is telling us that, at any fixed time $u$, for disjoint regions $A$ and $B$, the uncertainty is $\Delta\langle I_A(u)I_B(u)\rangle_{\rm out} = 0$.  However,  correlations emerge at different times. For example, for $N=4$, where $|\nu_i|=\ft 12$ for all $i$, we have
\be \gamma(u,u') =- \tan^2\left(\frac{\theta(u)-\theta(u')}{2}\right)\ .\ee
We can see the physics most clearly if we set $u=-T$ and $u'=+T$. In this case, 
\be \gamma(-T,+T) = -\frac{T^2}{L^2}\ .\ee
Similar growth is seen for all $N\geq 4$. This is telling us that the correlations between detections in disjoint regions $A$ and $B$ grow over time. This tallies with what we saw in Section \ref{manysec}, where there is a soft cloud of particles with frequency $\omega\ra 0$. The calculation above tells us that, perhaps unsurprisingly, these soft particles don't move in a single direction in three dimensions (as they're forced to in one dimension). Instead, they spread out in all directions, resulting in long-time correlations seen above.  

\subsection*{Long Time Correlations}

We can make this more precise. Above we have computed the correlations of currents. To get the correlation of charges, we simply need to integrate. It will prove interesting if we ask about correlations of the charge accumulated over some time $-T<u<+T$, which we write as
\be Q_A(T) = \int_{-T}^{+T} du\ I_A(u)\ .\label{qquick}\ee
We computed the one-point function in \eqn{qat}, where we saw that, if we wait long enough, we do indeed capture all the charge. 

\para
The integrated correlator \eqn{ii} gives
\be  \langle Q_A(T) Q_B(T)\rangle_{\rm out} &-& \langle Q_A\rangle_{\rm out} \langle Q_B\rangle_{\rm out} \nn\\ &=&   \langle Q_A(T) Q_B(T)\rangle_0  + \left[\frac{|A\cap B|}{4\pi} -  \frac{|A|\,|B|}{(4\pi)^2}\right]  {\cal R}(T) \label{qq}\ee 
%
where 
\be {\cal R}(T) = \int_{-T}^{+T}du\int_{-T}^{+T}du'\ \sum_{j=1}^N \frac{\sin^2(\nu_j[\theta(u)-\theta(u')])}{\pi^2 (u-u')^2}\ .\label{rt}\ee
The result \eqn{qq} includes a contribution from the vacuum fluctuations $\langle Q_A Q_B\rangle_0$. Here too there are divergences\footnote{The definition of $Q_A(T)$ given in \eqn{qquick} involves a sudden turning on and off of the detector at time $u=\pm T$. If this is smoothed out to some small ``turn-on'' time $\tau\ll T$, then fluctuations in any finite time measurement of the charge over a sphere of radius $r$, with $r\gg T \gg \eta$,  scale as $\langle Q(T)^2\rangle_0 \sim r^2/\tau^2$ \cite{bousso}.} which, physically, have the interpretation of quantum noise.


%
%
%
%
%
%

\para
If we focus on well-separated disjoint regions $A$ and $B$ with $A\cap B=0$, then the vacuum fluctuations, measured on a sphere of radius $r$, scale as 
\be \langle Q_A(T) Q_B(T) \rangle_0 \sim \frac{T^2}{r^2}\ee
If we sit at a distance $r\gg T$, we can ignore these. But we do get correlations  from the outgoing wavepacket and, for $A\cap B=0$, we have
\be  \langle Q_A(T) Q_B(T)\rangle_{\rm out} - \langle Q_A\rangle_{\rm out} \langle Q_B\rangle_{\rm out} =  -\frac{|A|\,|B|}{(4\pi)^2} {\cal R}(T) \ee
We can calculate the function ${\cal R}(T)$ in special cases. For $N=2$, we have 
\be  {\cal R}(T) = \left(\frac{2}{\pi}\tan^{-1}\left(\frac{T}{L}\right)\right)^2 \ \ \ {\rm for}\ \ N=2\ .\ee
This coincides with the disconnected piece $\langle Q_A(T)\rangle_{\rm out} \langle Q_B(T)\rangle_{\rm out}$ which we computed in \eqn{qat}. So in this case we have $ \langle Q_A(T) Q_B(T)\rangle_{\rm out} =0$ for disjoint regions. That's telling us, yet again, that just a single particle emerges from the scattering. Nothing funny is going on.

\para
For $N=4$, the integral can also be done analytically and this time we see something novel
\be {\cal R}(T) = \frac{4}{\pi^2} \log\left(1+\frac{T^2}{L^2}\right)\ \ \ {\rm for}\ \ N=4\ .\ee
Here we see the surprise: the correlations in  the  measured charge in disjoint regions grow stronger the longer you do the experiment. 
A similar  log growth is seen for all $N>4$,
\be {\cal R}(T) \ra \frac{2N}{\pi^2} \sin^2\left(\frac{2\pi}{N}\right) \log\left(\frac{T}{L}\right)\ \ \ {\rm as}\ \ T\ra \infty\ .\ee
 This is the physical manifestation of the scattering paradox: correlations  in the charge measured in disjoint regions grow over time. This growth can be tracked to the phase difference in the outgoing propagator \eqn{earlylate}: it is due to correlations in the pulse between very early and very late times. (If instead you choose to do the measurement over a time $T_-< u< T_+$ and send $T_+\ra \infty$ while keeping $T_-$ fixed, there is no log divergence.)

 %

\para
This behaviour would not be seen if there were a finite number of outgoing particles. It is the result of the infra-red structure of the pulse, with its collection of charged particle-anti-particle pairs with arbitrarily low energy.  This same  log growth is also seen in correlations between  flavour currents.

\subsection{Energy Correlators}

Next we turn to energy correlators. The energy flux observable is
\be {\cal E}_A(u) = \frac{i}{2} \sum_{i,l,l'} \int_A d^2\Omega\ Y_l^\dagger(\theta,\phi)Y_{l'}(\theta,\phi) :\left[\tchi^\dagger_{i,l}(u)\partial_u \tchi_{i,l'}(u) -  \partial_u\tchi_{il}^\dagger\,\tchi_{i,l'}(u)\right]:\ .\ee
The calculation of the two-point function has the same overall structure as the one above, but is slightly more intricate. Here we present only the final answer. The analog of \eqn{iaib} is 
\be \langle {\cal E}_A(u) {\cal E}_B(u')\rangle_{\rm out} &=& \langle {\cal E}_A(u) {\cal E}_B(u')\rangle_0 + \langle {\cal E}_A(u)\rangle_{\rm out}  \langle {\cal E}_B(u')\rangle_{\rm out}  \nn\\ 
&&\ \ +  \ \frac{|A\cap B|}{4\pi} \sum_j {\cal L}_j(u,u') + \frac{|A|\,|B|}{(4\pi)^2} \sum_j {\cal K}_j(u,u') \ee
where the functions ${\cal L}_j$ and ${\cal K}_j$ are given by
\be {\cal L}_j &=& -\frac{\sin^2(\nu_j[\theta(u) - \theta(u')])}{2\pi^2 (u-u')^4} - \frac{\nu_j^2}{2}\frac{|F(u)|^2 |F(u')|^2}{(u-u')^2} \cos(2\nu_j[\theta(u)-\theta(u')])\nn\\
{ \cal K}_j &=& \frac{\sin^2(\nu_j[\theta(u) - \theta(u')])}{2\pi^2 (u-u')^4} - \nu_j^2\frac{|F(u)|^2 |F(u')|^2}{(u-u')^2} \left[1-\frac{1}{2}\cos(2\nu_j[\theta(u)-\theta(u')])\right]  \nn\ee
They obey
\be {\cal L}_j(u,u') + {\cal K}_j(u,u') = -\nu_j^2 \frac{|F(u)|^2 |F(u')|^2}{(u-u')^2} \ .\ee
We can again check that in the cases $N=1$ and $N=2$, the excess two-point function (compared to the vacuum) is proportional to $|A\cap B|$, confirming that the outgoing state is  a single particle.  We can now evaluate the time integrals of this correlation function, but this time the integrals converge: unlike the charge, the fluctuations in the deposited energy do not grow with time.

\para
The simplest way to show this is to return to basics. We can define
\be E_A = \int_{-\infty}^{+\infty} du\ {\cal E}_A(u)\ .\ee
This has a particularly simple mode expansion.  We introduce creation and annihilation operators for each angular momentum mode -- $a_{i,l}(\omega)$ for the $i^{\rm th}$ species of particle and $b_{i,l}(\omega)$ for the $i^{\rm th}$ species of anti-particle with angular momentum $l=0,1,2,\ldots$ and energy $\omega$. Then we have
\be E_A = \int_0^\infty d\omega\ \omega \sum_{i,l,l'}\left[\int_A d^2\Omega\ Y^\dagger_l(\theta,\phi) Y_{l'}(\theta,\phi)\right]  \left(a_{i,l}^\dagger (\omega)a_{i,l'}(\omega) + b^\dagger_{i,l}(\omega) b_{i,l'}(\omega)\right)\ .\ \  \ \ \ee
A  straightforward calculation gives the two-point function of this operator as
\be \langle E_A E_B\rangle_{\rm out} = \frac{|A\cap B|}{4\pi} K^2 + \frac{|A|\,|B|}{(4\pi)^2} \left(\langle H^2\rangle_{\rm out} - K^2 \right)\ee
where $K^2$ is determined in terms of the number of particles $n(\omega)$ and anti-particles $\bar{n}(\omega)$ that we previously computed in \eqn{nomega} 
\be K^2 = \int_0^\infty d\omega\ \omega^2\sum_{i=1}^N\left[n_{\nu_i}(\omega) + \bar{n}_{\nu_i}(\omega)\right]\ .\ee
This should be viewed as a measure of the energy carried by a single excitation. We saw in Section \ref{manysec} that $n_\nu(\omega),\bar{n}_\nu(\omega) \sim 1/\omega$ as $\omega\ra 0$, but here we see explicitly that this infra-divergence doesn't affect the fluctuation of energies, since it is damped by the $\omega^2$ term in the definition of $K$. 

\para
We computed  $\langle H^2\rangle_{\rm out} = 1/2L^2$, as shown in \eqn{hh2}. For $N=1$ and $N=2$, we have $K^2=1/2L^2$, so that $\langle E_AE_B\rangle_{\rm out}$ is proportional to $|A\cap B|$.  This is again the statement that there is a single outgoing particle state in that case.  For  $N=4$, we can do the integrals exactly, and we have
\be K^2 = \frac{8}{3\pi^2L^2}\ .\ee
We then have
\be \langle E_A E_B\rangle_{\rm out} = \frac{1}{L^2} \left[ \frac{8}{3\pi^2} \frac{|A\cap B|}{4\pi} + \left(\frac{1}{2} - \frac{8}{3\pi^2}\right) 
\frac{|A|\,|B|}{(4\pi)^2}\right]\ .\label{355}\ee
We have correlations in the energy deposited in disjoint regions $A\cap B=0$. This is again telling us that the energy carried by the outgoing radiation is fragmented among different particles.  

\para
As we increase $N$, $K^2$ increases. We have the expansion
\be K^2 = \frac{1}{L^2} \left( \frac{1}{2} - \frac{7}{3N} + \frac{12-2\pi^2/3}{N^2} + \ldots \right)\ .\ee
This means that, at large $N$, the energy deposits become increasingly concentrated as if carried by a single hard excitation.  

\newpage
\section{What It Means}\label{summarysec}

We've gathered various pieces of evidence. Here we put them together.

\para
Recall the scattering paradox: in the s-wave, the ingoing particles sit in the representation ${\bf N}_{+1}$ of $SU(N)\times U(1)$, while the outgoing particles  sit in the representation ${\bf N}_{-1}$.  Yet scattering off the monopole preserves both $SU(N)$ and $U(1)$ symmetries, which means that the outgoing wavepacket carries the same representation ${\bf N}_{+1}$ as the ingoing particle.  For $N>2$, no finite collection of outgoing particles in the usual Fock space can reproduce these quantum numbers.

\para
We send in a one-particle state. In Section \ref{manysec} we showed, following \cite{yuji}, that this bounces back as a state with ill-defined particle number. We can count the number of particles $n(\omega)$ and anti-particles $\bar{n}(\omega)$ as a function of frequency $\omega$, and both exhibit a divergence $n(\omega), \bar{n}(\omega) \sim 1/\omega$ as $\omega\ra 0$. This is shown in Figure \ref{nnbarfig}.

\para
The ingoing wavepacket bounces back in the s-wave undeformed. But, unlike in $d=1+1$ dimensions, this wavepacket can fragment when measured, with particles moving in different directions. The fact that it does so shows up in the energy correlators. Take two regions $A$ and $B$ on the celestial sphere; even when the regions are disconnected, so $A\cap B=0$, there is a correlation in the energy detected in these regions, with $\langle E_A E_B\rangle_{\rm out} \neq 0$. This tells us that multiple particles are detected.

\subsection*{Hard Excitations}

So what is seen? There's nothing fishy about the outgoing state at any finite energy and the expectation is that, correspondingly, there's nothing fishy about any hard excitations that emerge from the scattering. They carry the standard quantum numbers of outgoing fermions meaning that they sit in the representation ${\bf N}_{-1}$. The remainder -- which is necessarily exotic -- is carried by the soft modes.

\para
Let's first discuss the hard excitations. To be specific, we will throw in a particle with flavour index $i=a$. Then what comes out is, naively, fractional particle numbers, where the  $i^{\rm th}$ comes out in fraction
\be \nu_i = \delta_{ia} - \frac{2}{N}\ .\ee
As we've seen, a better accounting is that the number of particles with flavour index $i$  and frequency $\omega$ is given by $n_{\nu_i}(\omega)$ defined in \eqn{nomega}. The number of anti-particles is  $\bar{n}_{\nu_i} (\omega)= n_{-\nu_i}(\omega)$. These functions don't quite translate into the probability of detecting a single particle of type $i$ with frequency $\omega$, because they count both single particle and multi-particle states. 

\begin{figure}[htb]
\begin{center}
\epsfxsize=6.2in\leavevmode\epsfbox{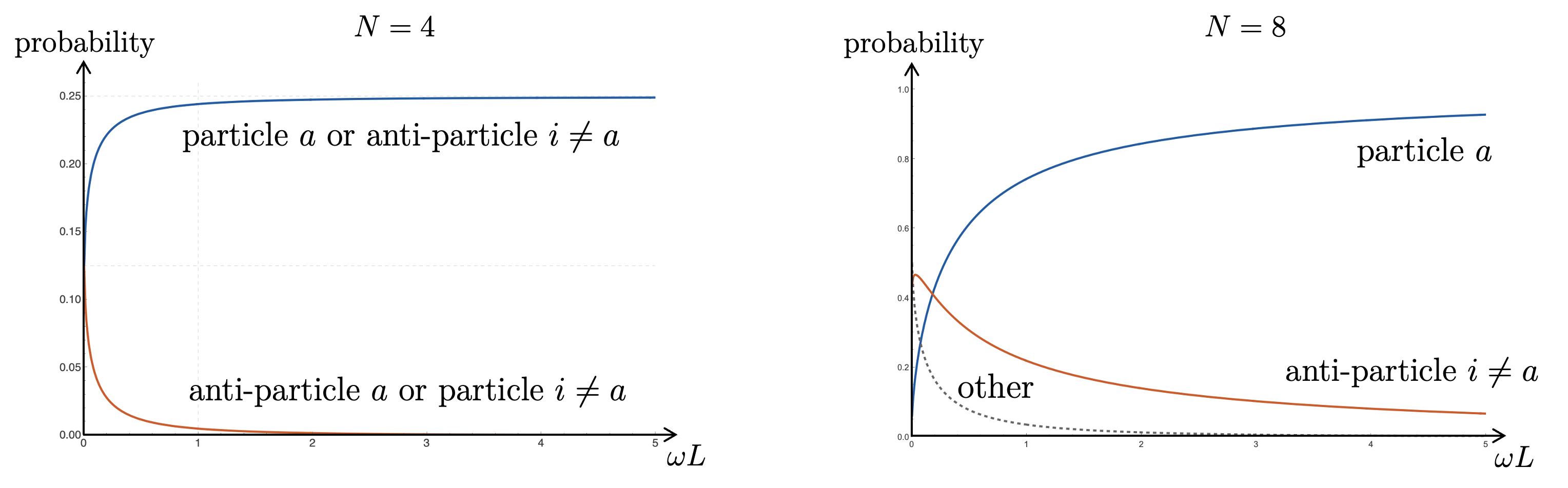}
\end{center}
\caption{The probability to detect different types of particles  as a function of frequency, for $N=4$ (on the left) and $N=8$ (on the right). In each case we throw in a particle of flavour $i=a$. Higher $N$ is qualitatively similar to the $N=8$ case.}
\label{detectfig}
\end{figure}

\para
Suppose, however, that we ask for the probability to detect particles in some small region $A$ with $|A|\ll 4\pi$. The probability to detect a single particle state is proportional to $|A|/4\pi$. The probability to detect two particles in this  region is proportional to $(|A|/4\pi)^2$. So for small regions, the probability to detect multiple particles through $|A|$ is much smaller than the probability to detect a single particle. Restricted to this context, we can then interpret $n_{\nu_i}(\omega)$ and $\bar{n}_{\nu_i}(\omega)$ as probabilities. 

\para
Specifically, suppose that we detect a single particle, in a small region $A$,  with energy $\omega$. The probabilities to see a particle of flavour $i$ or an anti-particle of flavour $i$ are
\be P(\mbox{particle $i$}) &=& \frac{n_{\nu_i}(\omega)}{\sum_j (n_{\nu_j}(\omega) + \bar{n}_{\nu_j}(\omega))} \nn\\ P(\mbox{anti-particle $i$}) &=& \frac{\bar{n}_{\nu_i}(\omega)}{\sum_j(n_{\nu_j}(\omega) + \bar{n}_{\nu_j}(\omega))}
\ee
These probabilities are shown in Figure \ref{detectfig}. The case $N=4$ is shown on the left. It is  special. If we throw in a particle with $a=1$, then the outgoing particles have
\be \nu = \left(\frac{1}{2},-\frac{1}{2}, -\frac{1}{2}, -\frac{1}{2}\right)\ .\ee
This means that the probability to detect a particle of flavour $i=1$ is the same as the probability to detect an anti-particle with flavour $i=2,3,4$. This is the upper-line in the graph. Meanwhile, the probability to detect an anti-particle of flavour $i=1$ is the same as the probability to detect a particle with flavour $i=2,3,4$. This is the lower line in the graph.  It is suppressed at high frequency by $1/(L\omega)^2$. 

\para
Suppose that we detect an outgoing particle with $i=1$. Then the $SU(4)$ flavour quantum number is accounted for, but there is a charge deficit of $+2$. Alternatively, if we detect an anti-particle of flavour $i=2,3,4$ then the charge is accounted for, but there is a deficit in the flavour quantum numbers, and the remaining outgoing state must be neutral and sits in the anti-symmetric ${\bf 6}$ of $SU(4)$.

\para
We might, of course, detect one of the suppressed particles instead (the lower line in Figure \ref{detectfig}). If we see a particle with flavour index $i=2,3,4$ then we leave behind an object with charge $+2$ in the adjoint ${\bf 15}$ of $SU(4)$. If we  detect an anti-particle with flavour $i=1$, we leave behind a neutral object in the symmetric ${\bf 10}$ of $SU(4)$.

\para
It may be that more than one hard particle emerges. But the scattering paradox means that, for any finite number of such particles, there will always be a deficit in either charge or flavour or both. This deficit must be carried by the soft cloud and we will see what that means below. Note that if we see a high energy particle, with $\omega L \gg 1$, the probabilities are
\be P(\mbox{particle $i=1$}) \ra \frac{1}{4} \ \ \ {\rm and}\ \ \ P(\mbox{anti-particle $i=2,3$ or 4)} \ra \frac{3}{4}\ .\ee
So it's more likely that the soft cloud has to carry flavour quantum numbers than charge quantum numbers in this case.  This is, perhaps, mildly surprising. The UV completion of our model replaces the boundary condition with the 't Hooft-Polyakov monopole. This comes with a rotor degree of freedom that, when excited, endows the monopole with electric charge \cite{julia}. A natural conjecture is that this rotor degree of freedom can be briefly, virtually, excited during scattering and then radiate the electric charge as a soft cloud. This appears not to be the only possibility, at least to the extent that the boundary condition correctly captures the physics of the 't Hooft-Polyakov monopole. 

\para
The story changes for larger $N$. As an example, the case of $N=8$ is shown on the right of Figure \ref{detectfig}. The relative probabilities to find a particle of flavour $i=a$ vs an anti-particle of flavour $i\neq a$ scale at high frequency as
\be  \frac{P(\mbox{particle $i=a$})}{P(\mbox{anti-particle $i\neq a$})} \sim \frac{\Gamma(2/N)^2}{(N-1)\Gamma(1-2/N)^2} (2L\omega)^{2-8/N}\ .\ee
A detection of a high energy excitation, with $\omega L \gg 1$, is now more likely to leave behind a charge deficit than a flavour deficit.

\subsection*{Soft Excitations}

The hard excitations are not exotic. But the essence of the scattering paradox is that they cannot account for the outgoing quantum numbers. That must be left to the soft excitations. And these manifest themselves in fluctuations.

\para
The fluctuations are those in the charge measured in an outgoing region $A$. In fact, we will have to ask a slightly more refined question. We leave our detector running for a time $-T < u < +T$ and measure the charge $Q_A(T)$ accumulated in this region over this time. (Recall that $u=t-r$ is the null time coordinate and the outgoing wavepacket is centered on $u=0$.)

\para
Even in the vacuum, there are  divergences in the fluctuations of $\langle Q_A(T)^2\rangle_0$. In the presence of an outgoing pulse, there are further contributions to these fluctuations, given by \eqn{qq}
\be \langle Q_A(T)^2\rangle_{\rm out} - \langle Q_A(T)^2\rangle_0 - \langle Q_A(T)\rangle_{\rm out}^2  =  \frac{|A|}{4\pi}\left(1-\frac{|A|}{4\pi}\right) {\cal R}(T)\ee
where ${\cal R}(T)$ is given in \eqn{rt}. At large times, it takes the form
\be {\cal R}(T) \ra \frac{2N}{\pi^2} \sin^2\left(\frac{2\pi}{N}\right) \log\left(\frac{T}{L}\right)\ \ \ {\rm as}\ \ T\ra \infty\ .\ee
The prefactor of $\frac{|A|}{4\pi} (1-\frac{|A|}{4\pi})$ is telling us that there is no ambiguity in the charge measured over the whole sphere. But in any given region, the fluctuations grow larger over time. This divergence can be traced to correlations between the front and back of the pulse. If you instead choose to do a measurement over a time  $T_-< u< T_+$ and send $T_+\ra \infty$ while keeping $T_-$ fixed, there is no log divergence.

\para
Moreover, the outgoing pulse exhibits correlations in charges measured in disjoint regions, $A\cap B=0$. Such correlations in the vacuum vanish if we measure the charge on a sphere of radius $r\gg T$. But, measured on the outgoing wavepacket, they take the form
\be  \langle Q_A(T) Q_B(T)\rangle_{\rm out} - \langle Q_A\rangle_{\rm out} \langle Q_B\rangle_{\rm out} = - \frac{|A|\,|B|}{(4\pi)^2}  \,{\cal R}(T)\ee
These too grow with time. The minus sign above is telling us that a positive charge through one region is most likely compensated by a negative charge through the other. 

\para
This, then, is what becomes of the exotic quantum numbers that can't be accounted for by any finite collection of outgoing particles. They live in the soft sector and get buried in the fluctuations. While a hermetic $4\pi$-detector sees the full charge, any attempt to reproduce this by adding together the contributions from small cells of area $A$ will be challenging. The charge measured in each cell fluctuates and these fluctuations grow over time. There are compensating fluctuations in other cells. Unlike the hard excitations, there is no reason to think these charge deposits  leave particle tracks. We're left with a situation analogous to the sign problem in numerical simulations: summing a large number of signed contributions, which must add up to the net charge.  That sum contains the  deficit -- whether in charge or in flavour quantum numbers -- that the finite energy particles cannot carry. The resolution to the monopole-fermion scattering paradox is, literally, lost in the quantum noise.

\subsection*{Acknowledgements}

I'm particularly grateful to David Kaplan, Ryan Plestid, and Sasha Zhiboedov for helpful conversations. I'd also like to thank Guillermo Arias-Tamargo, Philip Boyle Smith, Max Vel\'asquez Cotini Hutt, George Koundouros, Vazha Loladze, Rishi Mouland, and Take Okui for many discussions on monopole-fermion scattering over the past year. ChatGPT Astra was used extensively to help with calculations. The paper is human written. All em-dashes are mine.  This work was supported by STFC grant ST/X000664/1 and a Simons Investigator Award. 


\appendix

\section{Appendix:  Levitons -- A Condensed Matter Analogy}\label{levsec}

In this appendix, we describe a very simple condensed matter system where the same physics seen in fermion-monopole scattering arises. This is a  quantum wire, subjected to a time-dependent voltage pulse. The theory was developed in a seminal paper by Levitov, Lee, and Lesovik \cite{lev1}. 

\para
We start with massive fermions in $d=1+1$ dimensions. We take our fermions to be spinless, non-relativistic, and non-interacting.  The (unnormalised) single-particle energy eigenstate is\footnote{Because we're dealing with real physics here, we re-introduce $\hbar$ and $e$. It just feels right.} 
\be \psi(x) = e^{ikx} \ \ \ \Longrightarrow\ \ \ E=\frac{\hbar^2k^2}{2m}\ee
with $m$ the mass of the electron. (We could put our electrons in a finite box and normalise the wavefunctions, but it won't be necessary for what follows.) We now throw in many electrons. They fill states up to the Fermi energy $E_F$. The Fermi wavevector $k_F$ and the Fermi velocity $v_F$ are given by
\be E_F = \frac{\hbar^2 k_F^2}{2m} \ \ {\rm and}\ \ \ v_F = \frac{\hbar k_F}{m} \ .\ee
We now subject this Fermi sea to a voltage pulse. A pulse that is uniform in space just gives each state a time-dependent phase. But the physics is more interesting if we apply a spatially dependent pulse $V(x,t)$. The time-dependent Schr\"{o}dinger equation becomes
\be i\hbar\ppp{\psi}{t} = \left[-\frac{\hbar^2}{2m}\ppp{^2\psi}{x^2}  - e V(x,t)\right] \psi(x,t)\ee
with $-e$ the electron charge. 

\para
Here's what's going to happen. We will turn on a pulse where $V(x,t)$ varies spatially only over a small region near the origin. That means there is a spike of an electric field near the origin. This will excite  electrons which move to the right and holes which move to the left. The novelty is that both excitations will be accompanied by a cloud of electron-hole pairs and this cloud will generically have a soft divergence -- the number of electrons as a function of energy  $n(E)$, and the number of holes $\bar{n}(E)$,  will both scale as $n(E) \sim \bar{n}(E) \sim 1/(E-E_F)$.  This is the same divergence seen in monopole fermion scattering. Moreover, the charge of each propagating cloud will typically be fractional.

\para
Now for the calculation. We will turn the pulse on and off again over a time $L$. (The variable $L$ is an unusual choice for a time duration in non-relativistic quantum mechanics, but we choose it because it will coincide with the size of the wavepacket that we introduced in Section \ref{scattersec}.) Each filled energy eigenstate will evolve over this time, and settle down into a complicated combination of filled and unfilled states. We want to compute the number of excited electrons and holes in the system after the pulse has turned off.

\para
We take the duration of the pulse to obey
\be \frac{\hbar}{L} \ll E_F\ .\ee
This ensures that only those states near the Fermi surface will be affected by the pulse.  There are two Fermi points: right-movers at $k=k_F$ and left-movers at $k=-k_F$. Near the two Fermi points, we have
\be E \approx E_F \pm \hbar v_F (k\mp k_F)\ .\ee
We start by looking at what happens near  the right-moving Fermi point, $k=k_F$. We write the wavefunctions as
\be \psi(x,t) = e^{ik_Fx-iE_Ft/\hbar} A(x,t)\ .\ee
In the absence of the pulse, we have, for states close to the Fermi surface, $A(x,t) = e^{i(E-E_F) (x/v_F - t)/\hbar}$. We will assume that the envelope $A(x,t)$ varies more slowly than the $e^{ik_Fx}$ term, in which case the profile $A(x,t)$ obeys
\be i\hbar\left(\ppp{}{t} + v_F\ppp{}{x}\right) A(x,t) = -eV(x,t)A(x,t)\ .\label{aequation}\ee
Now we specify the pulse. We'll turn on the pulse only on the left-part of space, $x<0$, and only for some time period $\sim L$, which will give rise to a temporary electric field localised near the origin. At this stage, we're going to do something a little dodgy. We should vary the voltage over a distance scale $d$, with $k_F^{-1} \ll d \ll v_FL$, with the first of these inequalities necessary to justify the approximation that lead to \eqn{aequation}. Instead, we'll cheat a little and, for simplicity,  take the step function
\be V(x,t) = \left\{ \begin{array}{cc} V(t) \ \ &x<0 \\ 0 \ \ & x>0\end{array}\right.\ee
Ultimately, this shortcut won't change the long-time behaviour we're interested in. 

\para
For $x<0$, the equation \eqn{aequation} has the solution
\be A(x,t) = e^{i(E-E_F) (x/v_F - t)/\hbar} \, U(t)\ \ \ {\rm for}\  \ x<0\label{a9}\ee
where 
\be U(t) = \exp\left(\frac{ie}{\hbar}\int_{-\infty}^t dt' \ V(t')\right)\ .\ee
For $x>0$, we want a solution to \eqn{aequation} that patches continuously onto \eqn{a9}, so that $A(0^+,t) = A(0^-,t)$. That's achieved by 
\be A(x,t) =  e^{i(E-E_F) (x/v_F - t)/\hbar} \, U(t-x/v_F)\ \ \ {\rm for}\ \ x>0\ .\ee
This is telling us that the pulse at the origin leads to a disturbance that propagates to the right, as expected. 

\para
How should we think about this? Suppose that we multiplied the wavefunction $e^{i(E-E_F) (x/v_F - t)/\hbar}$ by the phase $e^{-i\Delta E (t-x/v_F)/\hbar}$. This would have the interpretation of changing the energy of the state from $E$ to $E+\Delta E$. Instead, we multiply by the phase $U(u)$, with $u=t-x/v_F$. We can Fourier expand this as
\be U(u) = \int d\epsilon\ \tilde{U}(\epsilon) e^{-i\epsilon u/\hbar}\ .\ee
The outgoing state is then
\be A(x,t) =  \int d\epsilon\ \tilde{U}(\epsilon) \,e^{-i(E+ \epsilon-E_F) u/ \hbar} \ \ \ {\rm for}\ \ x>0\ .\ee
So $\tilde{U}(\epsilon)$ is the amplitude to add energy $\epsilon$ to the state. 

\para
Before the pulse, all states with energies below $E_F$ were filled. The pulse gives some probability for each of those to be excited to an energy $E>E_F$, although this probability will be negligible for those  states deep within the Fermi sea. The total number of excited states of energy $E$ will then be
\be n(E) 
= \int_{E-E_F}^\infty d\epsilon\ |\tilde{U}(\epsilon)|^2\ .\ee
We can similarly compute the number of holes. This is 
\be \bar{n}(E) = \int_{E-E_F}^\infty d\epsilon\ |\tilde{U}(-\epsilon)|^2\ .\ee
To make further progress, we specify the time-dependence of the pulse. We will take it to have Lorentzian form
\be V(t) = -\frac{2\hbar\nu}{e} \frac{L}{t^2 + L^2} \ .\label{vt}\ee
The pulse is  characterised by the promised time scale $L$, together with $\nu \in \R$. The total time integral is given by
\be -\frac{e}{\hbar}\int_{-\infty}^\infty dt\ V(t) = 2\pi \nu\ .\ee
As we've seen, the physics depends on the integrated pulse
\be U(t) = \exp\left(\frac{ie}{\hbar}\int_{-\infty}^t dt' \ V(t')\right) = \exp\left(-2i\nu \int_{-\infty}^t dt' \ \frac{L}{t^{\prime\,2} + L^2}\right)\ .\ee
Notice, in particular, that for $\nu \notin Z$, this has a different phase on either side of the pulse, $U(-\infty)=1$ and $U(+\infty) = e^{-2\pi i \nu}$. As in the case of fermions scattering off monopoles, this will give rise to the curious fractional charge. 

\para
To see this, we can write the Fourier transform as
\be \tilde{U}(\epsilon) = \frac{1}{2\pi\hbar} \int dt\ e^{i\epsilon t/\hbar} U(t)  =  - \frac{1}{2\pi i\epsilon} \int dt\ e^{i\epsilon t/\hbar} \dot{U}(t)\ .\ee
First, let's calculate this for small $\epsilon$. Here we can view the phase $e^{i\epsilon t/\hbar}$ as effectively constant, so 
\be \tilde{U}(\epsilon) \sim \frac{1}{2\pi i \epsilon} \left[ U(\infty) - U(-\infty)\right] \ \ \ \Longrightarrow \ \ \ |\tilde{U}(\epsilon)|^2 \sim \frac{1}{\pi^2 \epsilon^2} \sin^2(\pi \nu)\ .\ee
Integrating then exhibits the soft cloud of electrons and holes at energies just above the Fermi energy
\be n(E) \sim \bar{n}(E) \sim \frac{\sin^2\pi \nu}{\pi^2 (E-E_F)}\ \ \ {\rm as}\ \ E\ra E_F\ .\ee
All of this mirrors exactly the story  for fermions scattering off monopoles. (See, for example, \eqn{nnbardiverge}.)

\para
We can also calculate the total number of excited electrons and holes. This is
\be N_e &=& \int_{E_F}^\infty dE\ n(E)  = \int_{E_F}^\infty dE \int_{E-E_F}^\infty d\epsilon\ |\tilde{U}(\epsilon)|^2 \nn\\ &=& \int_0^\infty d\epsilon \ |\tilde{U}(\epsilon)|^2 \int_{E_F}^{E_F+\epsilon} dE = \int_0^\infty d\epsilon\ \epsilon|\tilde{U}(\epsilon)|^2\ .\ee
Similarly, the total number of holes is
\be N_h = \int_0^\infty d\epsilon\ \epsilon |\tilde{U}(-\epsilon)|^2 \ .\ee
This gives the total charge counted from these excitations as 
\be N_e - N_h = \int_{-\infty}^\infty d\epsilon \ \epsilon |\tilde{U}(\epsilon)|^2 = \nu - \frac{\sin(2\pi\nu)}{2\pi}\ .\ee
As promised, the total charge is fractional for $\nu\notin \Z$.

\para
The story above was for the right-moving Fermi point, which sends a collection of excitations of charge $N_e-N_h$ propagating into the region $x>0$. Repeating this for the left-moving Fermi point at $k=-k_F$ gives a collection of excitations of charge $N_h-N_e$ propagating in the opposite direction, into the region $x<0$. 

\para
For condensed matter physicists, the interest is largely in the case $\nu\in \Z$, where the charge is integer. Here there is no soft cloud, and no electron-hole pair excitations accompanying the propagating particles. There are just $\nu$ electrons going one way, and $\nu$ holes going the other. This property doesn't happen for a general voltage pulse: it is special to the Lorentzian \eqn{vt}, and provides a method to inject clean, single electron states into a quantum conductor \cite{lev1,lev2,lev3}. These particles have been created experimentally and are given the name {\it levitons} \cite{dubois}. They provide a route to  gain control over quantum states, including tomography -- the  reconstruction of the quantum state through measurement \cite{cool}.

\para
There is also a discussion of {\it half-levitons} in the literature \cite{half}. These arise when $\nu=1/2$ and are equivalent to the  {\it semitons} seen in monopole-fermion scattering in the Standard Model, or the $N=4$ theory described in this paper.

\section{Appendix: The Particle Spectrum With Fixed Energy}\label{nwesec}

In Section \ref{scattersec}, we sent in, and got back out, a wavepacket of width $L$. This is not an energy eigenstate. Indeed, the expectation value of the energy is $\langle H\rangle = 1/2L$ which is comparable to its uncertainty, with $\langle H^2 \rangle= 1/2L^2$. 

\para
In this appendix, we show that one of the key properties -- the divergence $n(\omega)\sim 1/\omega$, is independent of the choice of wavepacket. We do this by computing the spectrum of particles $n(\omega|E)$ subject to the requirement that the total energy is $E$. 

\para
In Section \ref{scattersec}, we chose the ingoing state to be a one-particle state, with distribution of energies given by the wavepacket profile $f(\omega)= \ \sqrt{2L} e^{-L\omega}$ defined in \eqn{fomega}. This means that the probability density to measure energy $E=\omega$ is
\be P(E) = |f(E)|^2 = 2Le^{-2LE}\ .\label{pe}\ee
The evolution of the wavepacket is Hamiltonian, so this is also the probability distribution of energy of the outgoing wavepacket. But, as we saw in Section \ref{manysec} (and was previously derived in \cite{yuji}), the outgoing wavepacket does not have a well-defined particle number. This means that it contains many different states with the same energy $E$: a one-particle state, multi-particle states, and states with particle-anti-particle pairs. In Section \ref{manysec}, we counted the total number of particles and anti-particles with some fixed frequency $\omega$. Here we ask a slightly different question: suppose that you measure the outgoing state to have total energy $E$: what, then, is the distribution of particles and anti-particles with different frequencies.

\para
To answer this, we first decompose the outgoing state a
\be \ket{{\rm out}} = \int dE \ \sum_\alpha A_\alpha(E) \ket{E;\alpha}\ee
where $\alpha$ labels all the different states (one-particle, multi-particle, different flavours) that have total energy $E$. We know that the amplitudes must sum to 
\be P(E) = \sum_\alpha |A_\alpha(E)|^2\ee
where $P(E)$ takes the simple form \eqn{pe}.  Now define the projection operator onto states of energy $E$,
\be \Pi_E = \sum_\alpha\ket{E;\alpha}\bra{E;\alpha}\ee
If we measure $\ket{{\rm out}}$ to have total energy $E$ then it collapses to the state
\be \ket{{\rm out}_E} = \frac{\Pi_E\ket{{\rm out}}}{\sqrt{\bra{{\rm out}}\Pi_E\ket{{\rm out}}}}=\frac{\Pi_E\ket{{\rm out}}}{\sqrt{P(E)}} \ .\ee
Now, the number of particles of different frequencies in this state is
\be n(\omega|E) = \bra{{\rm out}_E} a^\dagger(\omega)a(\omega)\ket{{\rm out}_E}\ .\ee
We can relate this to the total number of particles $n(\omega)$ in $\ket{{\rm out}}$ by 
\be n(\omega) = \bra{{\rm out}} a^\dagger(\omega)a(\omega)\ket{{\rm out}} &=& \int\ dE\,dE'\ \bra{{\rm out}} \Pi_E a^\dagger(\omega) a(\omega)\Pi_{E'}\ket{{\rm out}}\nn\\ &=& \int dE\ P(E) \,n(\omega |E) \label{nwasnwe}\ee 
where we've used the completeness of the projection operators $\int dE\ \Pi_E=\mathds{1}$ and $\Pi_E\Pi_{E'} = \delta(E-E')$. 

\para
One advantage of computing $n(\omega|E)$ is that we can use it to construct the spectrum $n(\omega)$ for different choices of wavepackets. We simply use the formula \eqn{nwasnwe}, but with the probability distribution $P(E)$ determined by the wavepacket of choice. 

\para
We will now  compute $n(\omega|E)$. We start with our expression \eqn{nomega} for the particle number of a single species, labelled by $\nu$
\be n_\nu(\omega) = \int_\omega^\infty \frac{dk}{(2\pi)^2} |\widehat{U}_\nu(k)|^2\ee
where $\widehat{U}_\nu(k)$ is the Fourier transform of the phase
\be \widehat{U}_\nu (k) = \int_{-\infty}^{+\infty} du\ e^{ikv} \left(\frac{L-iu}{L+iu}\right)^\nu\ .\label{uft}\ee
The integral has a branch cut along the imaginary axis, starting at $u=iL$ and extending to infinity. We close the contour in the upper-half plane, running down, coming down just to the right of the imaginary axis, and up again just to the left, skipping around the branch cut. The difference between the two sides of the branch cut is $e^{i\pi\nu} - e^{-i\pi\nu} = 2i \sin(\pi\nu)$. We can then exchange the integral along the real axis in \eqn{uft} for an integral along the branch cut. If we write $u=iy$, then
\be \widehat{U}_\nu (k) = 2\sin(\pi\nu) \int_L^\infty dy\ e^{-ky} \left(\frac{y+L}{y-L}\right)^\nu\ .\ee
If we subsequently define $y=L+2Lt$, then
\be \widehat{U}_\nu (k) = 4L\sin(\pi \nu)  e^{-Lk} \int_0^\infty dt\ e^{-2Lkt}t^{-\nu} (1+t)^\nu\ .\ee
Substituting this into the expression $n_\nu(\omega)$, and doing the $\int dk$ integral, we get
\be n_\nu(\omega) = \frac{2L\sin^2(\pi\nu)}{\pi^2}\int_0^\infty dt\,dt' \ \frac{e^{-2L\omega(1+t+t')}}{1+t+t'}\left[\frac{(1+t)(1+t')}{tt'}\right]^\nu\ .\ee
Finally, we make the substitution $r=t+t'$, so that
\be n_\nu(\omega)  = \frac{2L\sin^2(\pi\nu)}{\pi^2}\int_0^\infty  dr\ \frac{e^{-2L\omega(1+r)}}{1+r}\int_0^r dt\ \left[\frac{1+t)(1+r-t)}{t(r-t)}\right]^\nu\ .\ee
This is now of the same form as \eqn{nwasnwe}, with the energy $E=\omega(1+r)$. We can write is as
\be n_\nu(\omega)  = \frac{2L\sin^2(\pi\nu)}{\pi^2} \int_\omega^E P(E) \int_0^{E/\omega-1} dy\ \left[\frac{(1+y)(E/\omega-y)}{y(E/\omega-1-y)}\right]^\nu\ .\ee
The uniqueness of the Laplace transform means that the integrand is $n(\omega|E)$. We have
\be n(\omega|E) = \frac{\sin^2(\pi\nu)}{\pi^2 E}\int_0^{E/\omega-1} dy\ \left[\frac{(1+y)(E/\omega-y)}{y(E/\omega-1-y)}\right]^\nu\ \ \ {\rm for}\  \ \omega\leq E\ .\label{nwe}\ee
and 
\be n(\omega|E) = 0\ \ \ {\rm for}\ \ \omega> E\ .\ee
The latter condition makes sense: if the total energy is $E$, we can't have any excitations with frequency $\omega> E$. Moreover, as expected the answer doesn't depend on the width of the initial wavepacket $L$. There are also a number of consistency checks. In particular, it's straightforward to show that in the case $\nu=1$, which is where we expect standard scattering, we retain a one-particle state, with
\be n(\omega|E) = \delta(\omega-E)\ \ \ {\rm when}\ \ \nu=1\ .\ee
For $\nu<1$, we have
\be n(\omega|E) \sim \frac{1}{E\Gamma(\nu)^2\Gamma(2-2\nu)}\left(\frac{E-\omega}{E}\right)^{1-2\nu}\ \ {\rm as}\ \ \omega \ra E\ee
We see that there is a divergence for $\nu>1/2$, while $n(\omega|E)\ra 0$ as $\omega\ra E$ for $\nu<1/2$. The case $\nu=1/2$ is borderline, with $n(\omega|E) \ra 1/\pi E$ as $\omega \ra E$. Meanwhile, the soft divergence remains for all $|\nu|<1$
\be n(\omega|E) \sim \frac{\sin^2(\pi\nu)}{\pi^2\omega}\ \ {\rm as}\ \ \omega\ra 0 \ .\ee
As in Section \ref{manysec}, the number of anti-particles is given by $\bar{n}_\nu(\omega|E) = n_{-\nu}(\omega|E)$.

\section{Appendix: A Mass Deformation}\label{masssec}

As we have seen, the non-standard quantum numbers that arise in monopole-fermion scattering are carried by soft modes. In this appendix we ask: what happens when those soft modes get a mass? This regularisation of the problem has been previously discussed in \cite{brennanplus,hook}.

\para
We take $N$ even and give all fermions an equal mass
\be {\cal L}_{\rm mass} = -m J_{ij} \tilde{\psi}_i\psi_j + {\rm h.c.} \ \  \ {\rm with}\ \ \ J= \left(\begin{array}{cc}\  \mathbb{0}  \ & \mathbb{1} \\ -\mathbb{1}\  & \mathbb{0}\end{array}\right)\ .\ee
This choice preserves an $Sp(N/2)$ subgroup of $SU(N)$ (in the convention that $Sp(1) = SU(2)$). At first glance, this seems to remove the scattering paradox completely: in the presence of a monopole, the ingoing s-wave fermions transform in ${\bf N}_{+1}$ and the outgoing fermions transform in ${\bf N}_{-1}$, but the ${\bf N}$ representation of $Sp(N/2)$ is pseudo-real. It is therefore possible for a ${\bf N}_{+1}$ particle to go in, and a $\overline{\bf N}_{+1}$ anti-particle to come out, preserving all quantum numbers. 

\para
This elastic scattering, in which a single particle goes in and a single particle comes out, is certainly possible. But that doesn't mean it happens dynamically. Indeed, if we take a wavepacket of width $L$, and average energy $\langle H \rangle = 1/2L \gg m$, then our previous analysis must hold for any high energy scattering. 

\para
To be concrete, we will focus on the case $N=4$. (Generalisations to higher $N$ are straightforward.) With massive fermions we have the option to include a theta angle, but we will set $\theta=0$. 

\para
We will take the mass to pair flavours $(13)$ and $(24)$. We then throw in a  particle with flavour index $i=1$. We saw in Section \ref{summarysec} that, for high energy scattering, there are four possibilities for the outgoing state, each occurring with probability $1/4$:  we could get a particle with flavour index $i=1$, or an anti-particle with flavour index $i=2,3,4$. We now look at each of these in turn.

\para
If a particle with flavour index $i=1$ emerges, that leaves a charge $+2$ unaccounted for. There are two options for what becomes of this. One is that there are two further outgoing anti-particles, carrying the charge. (These could be accompanied by further neutral particles.) The second is that, as shown by Hook and Ristow \cite{hook}, the fermions could  bind to the monopole, resulting in a dyon. These dyon states are surrounded by a fermion cloud of width $1/m$. The explicit calculations of \cite{hook}, using bosonisation, show that the would-be bound state for $N=4$ sits at threshold, but becomes truly bound for $N>4$, so in this case it seems that the deficit charge must be carried by two outgoing anti-particles (possibly with further particle-anti-particle pairs). 

\para
If, however, an anti-particle with flavour index $i=3$ emerges the story changes. This is paired with the ingoing particle in the mass matrix.  The charge is now accounted for, but not the flavour quantum numbers. The remainder was in the ${\bf 6}$ of $SU(4)$. Under $Sp(2)$, this decomposes as ${\bf 6}\ra {\bf 1} \oplus {\bf 5}$. The singlet allows for elastic scattering, while the ${\bf 5}$ must be accompanied by further particle-anti-particle pairs to account for the flavour symmetry. (The analysis of \cite{hook} finds no charge-neutral, flavoursome dyons able to soak up the quantum numbers.)  Finally, if an anti-particle with flavour index $i=2,4$ emerges then the charge is again accounted for, while the flavour index sits in the ${\bf 5}$ of $Sp(2)$ and must be accompanied by further particle-anti-particle pairs. In none of these cases is the scattering one-to-one.

\end{document}